\documentclass[11pt]{article}
\usepackage{color,helvet,times}
\usepackage{setspace} 
\usepackage{amsmath}
\usepackage{amssymb}
\usepackage{graphicx}
\usepackage{multirow}

\usepackage{natbib} \bibpunct{(}{)}{;}{author-year}{}{,}
\usepackage{amscd}
\usepackage{chemarrow}
\usepackage{verbatim}
\usepackage[normalem]{ulem}

\usepackage{geometry} 
\newcommand{\md}{d\kern-0.035cm\char39\kern-0.08cm}
\newcommand{\mL}{L\kern-0.15cm\char39}

\usepackage{calrsfs}

\renewcommand{\subsubsection}[1]{\paragraph{#1.}}

\def\ra{\ensuremath\rightarrow}
\def\ph{\ensuremath\varphi}

\begin{document}

\begin{center}
{\Large
\textbf\newline{Probabilistic models of individual and collective animal behavior} 
}
\newline
\\
Katar{\'\i}na  Bo{\md}ov\'a\textsuperscript{1*},
Gabriel J. Mitchell\textsuperscript{1},
Roy Harpaz\textsuperscript{2},
Elad Schneidman\textsuperscript{2},
Ga{\v s}per Tka{\v c}ik\textsuperscript{1}
\\
\bigskip
\textbf{$^1$} Institute of Science and Technology Austria, Am Campus 1, Klosterneuburg A-3400, Austria\\
\textbf{$^2$} Weizmann Institute of Science, Rehovot, Israel
\\
\bigskip

\end{center}

\begin{abstract}
Recent developments in automated  tracking allow uninterrupted, high-resolution recording of animal trajectories, sometimes coupled with the identification of stereotyped changes of body pose or other behaviors of interest. Analysis and interpretation of such data represents a challenge: the timing of animal behaviors may be stochastic and modulated by kinematic variables, by the interaction with the environment or with the conspecifics within the animal group, and dependent on internal cognitive or behavioral state of the individual. Existing models for collective motion typically fail to incorporate the discrete, stochastic, and internal-state-dependent aspects of behavior, while models focusing on individual animal behavior typically ignore the spatial aspects of the problem. Here we propose a probabilistic modeling framework to address this gap. Each animal can switch stochastically between different behavioral states, with each state resulting in a possibly different law of motion through space. Switching rates for behavioral transitions can depend in a very general way, which we seek to identify from data, on the effects of the environment as well as the interaction between the animals. We represent the switching dynamics as a Generalized Linear Model and show that: (i) forward simulation of multiple interacting animals is possible using a variant of the Gillespie's Stochastic Simulation Algorithm; (ii) formulated properly, the maximum likelihood inference of switching rate functions is tractably solvable by gradient descent; (iii) model selection can be used to identify factors that modulate behavioral state switching and to appropriately adjust model complexity to data. To illustrate our framework, we apply it to two synthetic models of animal motion and to real zebrafish tracking data.
\end{abstract}

* kbodova@ist.ac.at

\section{Introduction}
One of the most fascinating topics in interdisciplinary research is to understand the complexities of animal behavior in naturalistic settings. Many essential questions, explored previously either through direct observations or qualitatively using mathematical models, remain unanswered at the quantitative level that can connect to large scale data: How predictable is individual behavior on a moment-by-moment basis and what factors influence behavioral decisions? How can coordinated and collective motion realistically emerge  in groups of interacting animals? How do we include the existence of different ``cognitive'' or behavioral states into mathematical descriptions of animal motion?

Recent progress in automated recording techniques opened the door to addressing these questions, by making it possible to track single or multiple interacting animals for extended periods of time \cite{buhl2006,ballerini2008,gautrais2009,robinson2009,lukeman2010,nagy2010,katz2011,peshkin2015}. 
Traditional models of animal motion, constructed to favor simplicity and provide a qualitative match to the observed collective or aggregate (averaged) behaviors, are today being revisited and carefully fitted to extensive data to provide predictive, quantitative models~\cite{ballerini2008,lukeman2010,katz2011}.

In addition to revisiting existing models, the wealth of data also provides the motivation to devise new models and remove some of the previously-made restrictions and assumptions, or at least test rigorously whether they are essential. A central assumption that underlies many models of animal motion and behavior is that each animal continuously performs a universal computation to determine its next action or movement direction \cite{huth1992,vicsek1995,couzin2002}. Animals, however, often exhibit discrete stereotyped behaviors or behavioral switches that result in apparently very different modes of motion through space, much like ``kinematic proxies'' for different cognitive or behavioral states described by~\cite{Harpaz2017}.  
These individual animal states can be extracted from data using various approaches: low-dimensional reduction of the complex behavior \cite{stephens2008,berman2014,girdhar2015}, fragmentation of trajectories into segments based on the functionality of the space \cite{shemesh2013}, machine learning-based detection of behavior types \cite{branson2009} or other data mining methods \cite{nathan2012}. Moreover, the states can be defined also for the whole animal groups \cite{tunstrom2013}, which we do not consider here. 
The switches between the indidivual's behavioral modes seem to occur stochastically, but their rate could in principle be affected by many factors: by spatial preferences, by various environmental signals, by the motion of conspecifics within the group, etc. The question therefore arises about how to combine in a single tractable model, on the one hand, the kinematic description of individual's motion through space, and on the other, the existence of different discrete behaviors and behavioral states. If such  behavioral states are indeed present and important, then ignoring their existence---which happens in, e.g., classic zonal or force-field models---will enable us to learn only laws of behavior and motion that are ``averaged over'' various behavioral states. We would thus fail to capture the observed heterogeneity due to the stochastic and discrete nature of individual animal behaviors and, consequently, motion trajectories. 

The combination of stochastic state transitions with deterministic laws of motion through space distinguishes our model from the classical models where behavioral rules are assumed fixed in time. Traditionally, such rules are inferred by fitting to all available data, resulting in simulated trajectories which tend to be much smoother than real data. Our approach also differs from the population dynamics models capturing proportions of animals with different tasks or behaviors. For instance \cite{seely1995} and \cite{camazine1991} study behavioral states in honeybees when multiple food sources are present and distinguish between proportion of population feeding on different sources. Similarly, \cite{beckers1990}, \cite{detrain2006}, and \cite{sumpter2003} distinguish behavioral states in ant foraging due to either multiple food sources or due to recruiting of the resting ants. While these works describe different ``behavioral states'', they do so in terms of population averaged quantities, while our goal is to incorporate a set of different behaviors at the individual level and fit that to individual trajectory data.

We propose a class of probabilistic models that is sufficiently rich to capture a wide variety of complex individual and collective behaviors, and very flexible in how external factors modulate the focal animal's behavior. Despite this expressive power, the proposed models remain easy to simulate given the parameters, and support tractable maximum likelihood parameter inference from trajectory data. As described in the following section, these models are technically hybrid models that combine deterministic dynamics for animal motion with discrete behavioral states; stochastic transitions between these states are described in a Generalized Linear Model framework. 

This paper is structured as follows. The next section introduces the rationale behind the modeling approach that we propose, specifies the model mathematically, and discusses the forward (simulation) and inverse (parameter inference) approaches. The second half of the paper focuses on three examples: two synthetic toy models of animal behavior (ant motion, bacterial chemotaxis) to showcase and validate the inference as well as illustrate an interesting extension to coarse-grained behavioral states, and one real data example (tracked zebrafish data) to illustrate inference without prior knowledge of laws of motion, as well as model selection to identify explanatory variables that affect behavioral state switching.

\section{Methods and models}  \label{sec_Methods}

The main goal of this work is to develop a probabilistic approach for understanding the behavior and motion of animals, either in an individual setting or as a group in which the animals can interact. We construe the ``behavior'' very broadly, as spanning the range from the occurrence of stereotyped events that can be identified by the body pose, specific interactions with an environment, vocalization production, etc., to internal behavioral states that manifest with the animal adopting a particular center-of-mass law of motion through space. We assume that each animal switches between the possible behaviors in a stochastic fashion, with rates that depend, in a very general way that we seek to identify, on the animal's current behavioral state, its location, and the location of other animals in a group.

\begin{figure}[h]
	\centering
	\includegraphics[width=0.8\textwidth]{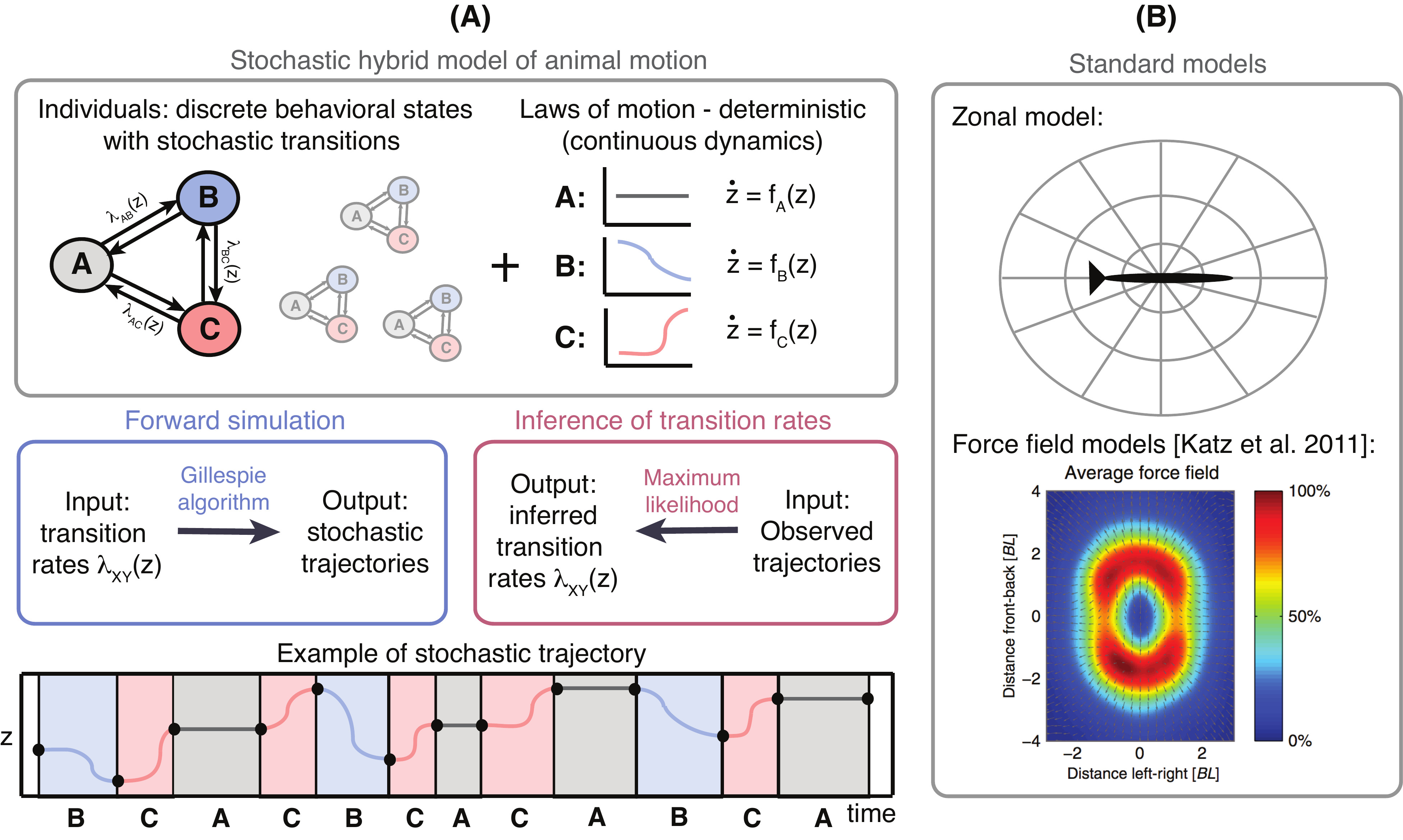}
	\caption{{\bf Models of animal motion and inference of model parameters.}
	{\bf (A)} The hybrid model consists of stochastic switching between the behavioral states (denoted as A, B, C) combined with
	the deterministic laws of motion for each state. When the behavioral states, 
	laws of motion for each state, and the transition rates among them 
	are known, the sample trajectories can be obtained by a stochastic simulation algorithm (i.e., the Gillespie algorithm). 
	However, if the measured animal trajectories can be classified into the known behavioral states 
	but the laws of motion and transition rates
	are unknown, this leads to an inference problem for the unknown transition rates.
	{\bf (B)} Standard models, e.g.,  
	the zonal models \cite{lukeman2010} and the force-field models \cite{katz2011},
	do not capture discrete stereotyped behaviors and assume a single, universal computation carried out
	by each animal at every instant to determine the animal's motion. The bottom subpanel of (B) is reproduced from 
	\cite{katz2011}, supporting information. 
	 }
	 \label{fig_BigPicture}
\end{figure}

The expressive power of our model stems from the combination of its stochastic and deterministic components, as schematized in Fig~\ref{fig_BigPicture}A and described in detail in Section~\ref{ssec_models}. Formally, our model belongs to a class of hybrid models \cite{pola2003,davis1993,ghosh1997,hu2000} with stochastic discrete states and deterministic continuous dynamics, typically called Piecewise-Deterministic Markov Processes (PDMP). While switching between behavioral states is stochastic, within each behavioral state the animal follows a certain law of motion that governs its kinematics (e.g., its center-of-mass trajectory), as specified by a system of ordinary differential equations. This is a powerful generalization of the traditional models, schematized for comparison in Fig~\ref{fig_BigPicture}B: in either zonal or force field models, the focal animal always follows a single, universal law of motion that cannot depend on the behavioral state. Similarly, frameworks that model switching between behavioral states of individual animals often ignore the fact that this behavior plays out in physical space and is modulated by the ever-changing spatial relationship between the focal animal, its conspecifics, and its environment. Our model brings together the analysis of discrete behavioral states and motion through space in a single probabilistic framework.

Given our very general notion of what constitutes a ``behavior,'' there exists no method for automatically detecting and annotating such behaviors in a way that is independent of the animal species, experimental setup, and the biological question being asked. For example, when describing ants, the most coarse-grained and easily observable behaviors might be the modes of locomotion (such as stopping and moving forward); however, in  other experimental setups we might want to sidestep motion details and instead focus on anntenation and grooming behaviors triggered by  exposure to specific pathogens. In the absence of generic algorithms for behavioral annotation, we start here by assuming that we are already given individual animal motion trajectories that have been annotated with behaviors of interest; this input data is obtained from raw observations in a problem-specific way that we do not address here. Starting with such input data, our framework will then proceed to identify how the animals switch between the identified behaviors, using a generic, problem-independent inference procedure. In this way, the complex problem of modeling animal behavior and motion can be broken down into manageable pieces, some of which  admit generic solutions and thus need to be studied only once.

Even though a generic mechanism for identifying behaviors is currently out of reach, we use several examples of animal motion in Section~\ref{sec_examples}
to show how such identification can be attempted in practice for specific cases where kinematic proxies exist for behavioral states. Our three examples include the case of simulated ant motion and bacterial chemotaxis, and the case of real (tracked) zebrafish data. Because our modeling framework is probabilistic, it naturally captures the stochastic nature of animal decision-making and can tolerate a degree of noise in the data. As we show in
Section~\ref{ssec_simulation}, 
the structure of our model permits straightforward and tractable forward simulation of stochastic animal trajectories (of motion as well as behavioral states) in a group, using a variant of Gillespie's stocahstic simulation algorithm (SSA)~\cite{gillespie1977}. Conversly, given the data, we can use the standard statistical tools to solve the inverse problem of model inference and model comparison to identify environmental and group factors that modulate behavioral transitions. Importantly, since our model is a derivative of well-studied Generalized Linear Models (GLMs) in statistics and neuroscience, we can guarantee  that maximum likelihood inference is convex. This means that for identifiable models there is a single set of best fitting parameters for the dependence of behavioral switching rates for each animal, which can be solved sequentially, animal-by-animal, by standard gradient descent, as we show in Section~\ref{ssec_inference}. 
Taken together, these methodological properties of our framework ensure that it remains practically applicable for inference and simulation even while it maintains its large flexibility and expressive power.

\subsection{Model of animal motion} \label{ssec_models}
The model's main components are schematically displayed in Fig~\ref{fig_BigPicture}. We consider a population of animals of size $N$ where the $n$-th individual is characterized by its position $\mathbf z_n = (z^1, \dots, z^d)$ in $d$ dimensions. This position can be complemented by other physically relevant kinematic measurements, for instance velocity, orientation, etc; for simplicity, we will nevertheless refer to $\mathbf{z}$ simply as ``position.'' In addition, each individual has an internal behavioral state, $x_n$, chosen from a set of available states,  $\{ 1, \dots, S \}$. Each individual at every time follows a deterministic law of motion which describes the rate of change of the position (velocity, etc.) depending on the current position and state of the animal
\begin{eqnarray}
	\frac{d \mathbf z_n}{dt} = f(\mathbf z,x_n,t)=f_{x_n}(\mathbf z,t), \quad n\in\{1,\dots,N\}\,. \label{eq_ODE}
\end{eqnarray}
Stochasticity of the motion arises due to random switching between the internal states with the transition rates $\lambda(x_n\rightarrow x_n',\mathbf z_1,\dots,\mathbf z_N)$. These transition rates depend not only on the states relevant to this transition, but also on the position of (possibly all)  other animals, and on other important dynamic or environmental factors, captured by $\mathbf z$. 
The central assumption of our model concerns the mathematical form through which the switching rates $\lambda$, characterized by a set of parameters $\alpha_i^{(1)}, \alpha_{ij}^{(2)}$, depend on the internal states and positions:
\begin{eqnarray}
	&&\lambda(x_n \ra x_n',\mathbf z_1,\dots,\mathbf z_N) = \nonumber \\
	 &&g \left( \sum_i \alpha^{(1)}_i(x_n \ra x_n') \ph_i(\mathbf z_n) + \sum_{m\neq n} \sum_{i j} \alpha^{(2)}_{i j}(x_n \ra x_n') \Psi_{i j}(\mathbf z_n,\mathbf z_{m})
	  \right) \,.\label{eq_rates}
\end{eqnarray}
Animal positions, $\mathbf z$, in this relationship serve as a ``stimulus'' that drives the changes of state by modulating transition rates $\lambda$. The dependence on $\mathbf z$ is made tractable by choosing a particular representation of positions, $\ph_i$ and $\Psi_{ij}$, which essentially discretizes the space of positions, as discussed below. For example, in Eq~(\ref{eq_rates}), representations $\ph_i$ and $\Psi_{ij}$ parametrize the dependence of switching rates on position coordinates of individual animals (through the coefficients $\alpha_i^{(1)}$), and on pairwise spatial relationships between the animals (through $\alpha_{ij}^{(2)}$), excluding interactions of more than two individuals. Formally,  rates $\lambda$ are the instantaneous rates of an inhomogeneous Poisson point processes describing behavioral state changes that are approximated as infinitely fast. The dependence of the rates on  position and state variables has the Generalized Linear Model form~\cite{nelder1972,mccullagh1989}: the transition rates are specified by a nonlinear function, $g$, acting on a summation over representations of position ($\ph(\mathbf{z}_n)$ and $\Psi(\mathbf{z}_n, \mathbf{z}_m)$)  that is linear in sought-for parameters $\alpha$. In statistical literature, the inverse of the nonlinear function $g$ is also known as the ``link function.'' In sensory and motor neuroscience, similar models have been used to analyze the dependence of Poisson spiking of neurons on the stimulus and output of other, simultaneously recorded, neurons in a neural circuit~\cite{pillow2008}.

Applied to animal behavior, it is easiest to understand the model of Eq~(\ref{eq_rates}) when the animals are not interacting. In this case, $\alpha_{i j}^{(2)}=0$, the second term in the argument to $g(\cdot)$ in Eq~(\ref{eq_rates}) vanishes, and behavioral transition rates can be decomposed into coefficients $\alpha^{(1)}$ (that depend on the kind of behavioral transition) that are multiplicatively modulated by some representation of animal's position, $\ph_i(\mathbf{z}_n)$. The choice of representations, $\ph$, is fixed prior to model inference, and can provide a lot of flexibility as well as regularization of the parameters $\alpha^{(1)}$. The parameters $\alpha^{(1)}$ are inferred from data using maximum likelihood inference. If the animals interact in a pairwise fashion, the second term, parametrized by $\alpha^{(2)}$, and also inferred by maximum likelihood, can be added to the argument of the nonlinear function. A typical example for the choice of $\Psi$ in this case could be a function that depends on the distance between the center-of-masses of interacting animals $n$ and $m$. 

\begin{figure}[h]
	\centering
	\includegraphics[width=1\textwidth]{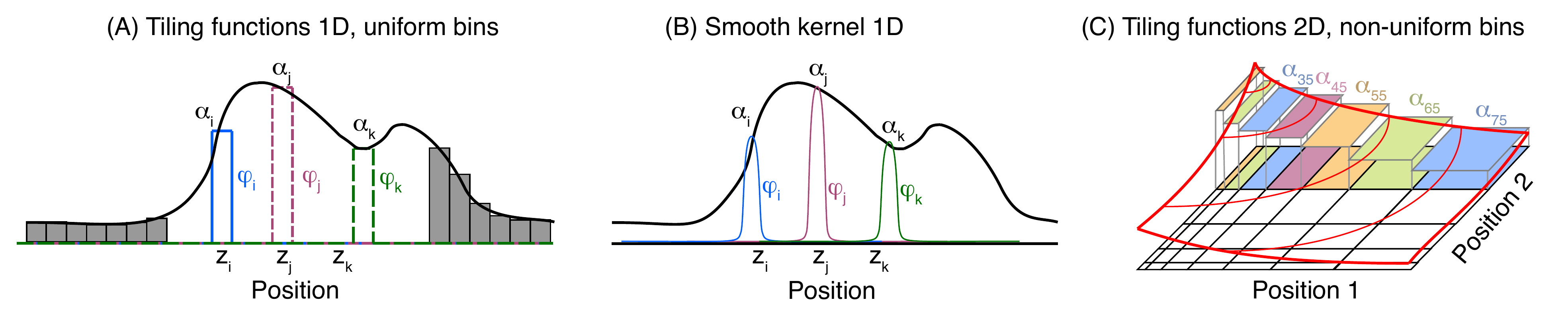}
	\caption{{\bf Dependence of behavioral transition rates on $\mathbf z$ using different representations of position.}
	{\bf (A)} The simplest way to approximate a transition function without any further assumptions on its shape 
	is to discretize it using equidistant binning. This represents the rates using a linear combination of tiling functions $\varphi_i$, 
	which have value 1 in a narrow region of the position space and 0 otherwise. 
	The multiplicative constants $\alpha_i$, that are inferred in our approach, then determine the shape of the transition rate function.
	We choose non-overlaping tiling regions which cover the whole position space. 
	Smaller size of the tiling regions leads to a more accurate approximation of the transition rate dependence 
	on the position variables but requires more data for inference.
	{\bf (B)} In this example, the rates are expanded into a linear combination of Gaussian bump functions that tile the domain. This 
	enforces the smoothness of the rate at the spatial scale that corresponds to the width of the Gaussians. 
	{\bf (C)} An example of tiling functions for representing the rate on the 2D domain. In contrast to (A), here the bin sizes are not chosen uniformly.
	}
	 \label{fig_tiling}
\end{figure}

There exist many possibilities for the representation of the kinematic variables $\mathbf{z}$ (i.e., for the functions $\ph$ and $\Psi$) and the choice between them is problem-specific. 
In the absence of any prior information about how animal position could affect the transition rates, we choose here the tiling functions, schematized in Fig~\ref{fig_tiling}A and C. 
Tiling functions represent nothing else but some particular discretization of the continuous domain, e.g.,  $\mathbf z \in \mathbb{R}^d$, into bins; note that the bins can be non-uniform, as in Fig~\ref{fig_tiling}C. The choice of tiling functions is particularly useful when the positions of individuals can be measured with limited resolution. With the domain discretized into bins, the dependence of switching rates on position in Eq~(\ref{eq_rates}) is simply represented point-wise, by values of $\alpha$ specified bin-by-bin. Representation functions $\ph$ and $\Psi$ can therefore be thought of as a ``basis'' in which we expand the dependence of switching rates on positions, with $\alpha$ being the coefficients of the expansion in the chosen basis. 
\begin{table}
\caption{Abbreviations and notation.}
 \label{tab_notation}
\begin{tabular}{|l|p{10cm}|}
	\hline
	$N$ & number of individuals\\
	$d$ & dimension of the space of motion\\
	$\mathbf z_n = (z_n^1,\dots, z_n^d)$ & kinematic variables (e.g., position) of the $n$-th individual\\
	$\mathbf z = (\mathbf z_1,\dots, \mathbf z_N)$ & kinematic variables of all the individuals \\
	$S$ & number of behavioral states\\
	$X = (1,\dots, S)$ & space of available behavioral states\\
	$x_n \in X$ & internal state of the $n$-th individual\\
	$\mathbf x \in X^N$ & internal state of all individuals\\
	$L(\cdot)$ & log-likelihood function\\
	$\boldsymbol \alpha^{(1)}$ & generalized linear model parameters of the first order\\
	$\boldsymbol \alpha^{(2)}$ & GLM parameters of the second order (interaction)\\
	$k' \in K_S$ & time bin indices (with corresponding times $t_{k'}$) at which the trajectories were sampled \\
	$k \in K_T$ & time bin indices (with corresponding times $t_k$) when state transitions occurred \\
	\hline
\end{tabular}
\end{table}

The benefit of the tiling functions is the simplicity of their interpretation and implementation. In what follows we will see that with the tiling functions the sufficient statistics for maximum likelihood inference are simply contingency tables (or histograms), counting the number of behavioral transition events of each type occurring in every bin. The drawback of the tiling functions is the curse of dimensionality and the absence of smoothness regularization. If dimensionality of the kinematic space, $d$, is larger than typically 2 or 3, naive discretization along each dimension will result in very high dimensional histograms that will be poorly sampled, and hence the transition rates at many values of the kinematic variables will be unconstrained by the data. Tiling functions also permit the rates to depend on the position in an arbitrary, possibly non-smooth way. If, however, smoothness is expected, other basis choices (for instance, Gaussian bumps that tile the domain as in Fig~\ref{fig_tiling}B) may offer better generalization performance even with the basis of lower dimensionality. Ultimately, any choice of position representation is possible, with the inference remaining convex and tractable so long as: {\bf (i)} the dependence of the argument of nonlinear function, $g$, is linear in parameters $\alpha$, to be inferred; {\bf (ii)} the link function, $g^{-1}$, is chosen to be convex and log-concave~\cite{paninski2004}. A convenient choice satisfying these conditions that we adapt here for simplicity is $g(\cdot) = \exp(\cdot)$, which also implies that covariates in Eq~(\ref{eq_rates}), which are additive as arguments to $g$, have multiplicative effects on the switching rates $\lambda$.

The stochastic process described in Eq~\eqref{eq_rates} is a non-homogeneous continuous-time Markov chain, since the transition rates of the $n$-th individual only depend on its current state $x_n$ and on the external time-dependent input $\mathbf z_1,\dots,\mathbf z_N$ of all individuals. Therefore the probability density of occupying state $\mathbf x = (x_1,\dots,x_N)$ and $\mathbf z = (\mathbf z_1,\dots,\mathbf z_N)$ obeys a Master equation
\begin{eqnarray}
	\frac{\partial p(\mathbf x,\mathbf z,t)}{\partial t} 
	&=& \sum_n \sum_{ \mathbf x' \in \mathcal{N}_n(\mathbf z)} 
	\left [ \lambda(\mathbf x'_n \ra \mathbf x_n,\mathbf z) p(\mathbf x',\mathbf z,t) 
	 - \lambda( \mathbf x_n \ra  \mathbf x'_n,\mathbf z) p(\mathbf x,\mathbf z,t)  \right ]\nonumber\\
	 &&- \nabla_{\mathbf z} \cdot (f(\mathbf z,\mathbf x,t)p(\mathbf x,\mathbf z,t))\,,\label{eq_CME}
\end{eqnarray}
where $\mathcal{N}_n(\mathbf z)$ for $n\in \{1,\dots,N\}$ are disjoint sets of all behavioral states that can be reached from the state $\mathbf x$, differing from the state $\mathbf x$ only in the $n$-th component. While the first two components in Eq~\eqref{eq_CME} represent changes in the behavioral state of an individual, the last term accounts for the continually changing position of the individuals. The crucial assumption is that the law of motion is deterministic, i.e., all stochastic influences can be characterized as random transitions between suitably defined behavioral states. The total number of states $\mathbf x \in \{1, \dots ,S\}^N$ may be large but discrete and the position space is continuous.

\subsection{Forward problem: stochastic simulation} \label{ssec_simulation}

Stochastic simulations of the animal motion are performed by a variant of the Gillespie algorithm \cite{gillespie1977}, which is an exact algorithm generating realizations of the process given by Eq~\eqref{eq_CME}. The system starts in the initial state $(t_0,\mathbf x(t_0),\mathbf z(t_0))$, characterized by the initial time and the states and positions of all individuals. Then it undergoes a series of transitions through the states $\{t_k,\mathbf x(t_k),\mathbf z(t_k)\}$ where $t_k$ are the transition times, indexed by $k$. Between these transition times, the states do not change but each individual follows continuous laws of motion determined by the state which the individual currently resides in.

The instantaneous transition rate is the sum of the individual rates out of the current state $\mathbf x$ through all possible transitions $\mathbf x \ra \mathbf x'$ which lead to a change of the state:
\begin{eqnarray}
	\lambda(\mathbf x,\mathbf z,t) 
	=  \sum_n \sum_{ \mathbf x' \in \mathcal{N}_n(\mathbf z)}  \lambda(\mathbf x_n \ra  \mathbf x'_n,\mathbf z)\,.
\end{eqnarray}
So if no transition occurs in a time period $[t_0,t_f]$, the rate accumulates to  
\begin{eqnarray}
	\Lambda(\mathbf x,\mathbf z(t_0),t_f,t_0) 
	= \int_{t_0}^{t_f} \lambda(\mathbf x,\mathbf z(t'),t') dt'\,,
\end{eqnarray}
where $\mathbf z(t)$ follows a deterministic dynamics in Eq~\eqref{eq_ODE}.
Because the stochastic process $(\mathbf x,\mathbf z)$ is Markovian, the waiting time until the next transition is exponentially distributed with rate $\Lambda$:
\begin{eqnarray}
	p(t_f | \mathbf x,\mathbf z(t_0),t_0) 
	= \lambda(\mathbf x,\mathbf z,t_f) e^{-\Lambda(\mathbf x,\mathbf z(t_0),t_f,t_0)}\,. \label{eq_exp}
\end{eqnarray}
Each transition can be realized by first choosing the individual that changes its state and then selecting the particular transition. This is true because of the following factorization of the transition probability:
\begin{eqnarray}
	p(\mathbf x' | \mathbf x, \mathbf z(t_f),t_f) 
	= p(n | \mathbf x, \mathbf z(t_f),t_f) p(\mathbf x'_n| \mathbf x_n, \mathbf z(t_f),t_f)\,.
\end{eqnarray}
The stochastic simulation algorithm uses the distribution of waiting times until the next transition given by Eq~\eqref{eq_exp} and the probabilistic rule for picking this transition, and can be summarized in the following steps:
\begin{itemize}
	\item[1.] Initialize: $\mathbf x_n(t_0),\mathbf z_n(t_0),t_0$, $k=1$.
	\item[2.] Compute rate functions: $\lambda(\mathbf x_n \ra \mathbf x_n',\mathbf z_n,t)$.
	\item[3.] Generate random numbers: $r_1 \sim \text{Exp}(1)$, $r_2,r_3\sim U([0,1])$.
	\item[4.] Compute time $t_{k}$ to the next transition: $\Lambda(\mathbf x,\mathbf z(t_{k-1}),t_k,t_{k-1}) =r_1$  where $\mathbf z$ solves $\mathbf z' = f(\mathbf z,\mathbf x,t)$.
	\item[5.] Choose an individual $n$ randomly (using $r_2$) with weights: $\lambda(\mathbf x_n,\mathbf z,t_k)/\lambda(\mathbf x,\mathbf z,t_k)$.
	\item[6.] Choose the transition $x'_n$ randomly (using $r_3$) with weights: $\lambda(\mathbf x_n \ra \mathbf x'_n,\mathbf z,t_k)/\lambda(\mathbf x,\mathbf z,t_k)$.
	\item[7.] Update $\mathbf x_n(t_k), \mathbf z_n(t_k), t_k$, set $k = k+1$ and repeat from 2 until the terminal condition.
\end{itemize}

\subsection{Inverse problem: parameter inference} \label{ssec_inference}

The inference uses experimental data and their annotation with the behavioral states to obtain the transition rates between these states. We denote the data as $\mathcal{D} = \{ t_{k'},\mathbf x(t_{k'}) ,\mathbf z(t_{k'}) \}$ where index $k'\in K_S$ is used to specify sampling times $t_{k'}$. We furthermore say that a transition $\mathbf x_{k-1}\ra \mathbf x_k$ occurred at time $t_k$, $k\in K_T$ if $\mathbf x_{k-1}\neq \mathbf x_k$ (where $K_T$ denotes the set of all transition time indices). In principle, if deterministic laws of motion were known in advance, it would be sufficient to record just the times at which a transition occurs, since the rest of the trajectory can be calculated from the known laws of motion. In practice, however, the laws of motion $f(\mathbf z, x_n,t)$ are often not known and the system has to be sampled finely even between the transition times. This is the regime we assume here. Thus, in case of multiple animal recordings, the two arrays $\mathbf x_{k-1}, \mathbf x_k$ differ at most in one component, corresponding to a single animal changing its state at any given transition time. 

For simplicity, we assume that all animals are identical, i.e., that the parameters $\alpha$ in Eq~(\ref{eq_rates}) are equal among the individuals. This assumption is not necessary for the tractability of the inference, and if data is sufficient, parameters can be inferred on a per-animal basis to ask questions about animal-to-animal heterogeneity. 

We  express the probability of observing a sequence of all transitions $\{ t_k,\mathbf x(t_k),\mathbf z(t_k)  \}$, $k\in K_T$ corresponding to the parameters $\boldsymbol \alpha$ as
\begin{eqnarray}
	p(\{ \mathbf x \} | \{ \mathbf z\}, \boldsymbol \alpha) &=& 
	\prod_{k\in K_T} p(\mathbf x(t_k), t_k | \mathbf x(t_{k-1}),t_{k-1}) \nonumber\\
	&=& \prod_{k\in K_T} p(\mathbf x(t_k) | \mathbf x(t_{k-1}),t_k) p(t_k | \mathbf x(t_{k-1}),t_{k-1})\,. \label{eq_path}
\end{eqnarray}
The probabilities are conditioned on the observed values $\mathbf z$ and parameters $\boldsymbol \alpha$ in Eq~\eqref{eq_rates}. We next construct a log-likelihood function $L(\boldsymbol \alpha|\mathcal{D})$ given the data $\mathcal{D}$. For a single individual this becomes 
\begin{eqnarray}
	L(\boldsymbol \alpha|\mathcal{D}) &=& \ln p(\{\mathbf x\}|\{\mathbf z\},\boldsymbol \alpha) + \ln p(\boldsymbol \alpha) + C 
	\nonumber \\
	&=& \sum_{k\in K_T} \left( 
	\ln p(\mathbf x(t_k) | \mathbf x(t_{k-1}),t_k) +  \ln p(t_k | \mathbf x(t_{k-1}),t_{k-1}) 
	\right)  + \ln p(\boldsymbol \alpha) +C \nonumber \\
	&=& \sum_{k\in K_T} \left( 
	\ln \frac{\lambda(\mathbf x(t_{k-1}) \ra \mathbf x(t_k), \mathbf z(t_k) | \boldsymbol \alpha )}
	{ \sum_{x'}\lambda(\mathbf x(t_{k-1}) \ra \mathbf x', \mathbf z(t_k) | \boldsymbol \alpha )} 
	+  \ln \frac{  \sum_{x'}\lambda(\mathbf x(t_{k-1}) \ra \mathbf x', \mathbf z(t_k) | \boldsymbol \alpha ) }
	{ e^{  \sum_{x'} \int_{t_{k-1}}^{t_k} \lambda(\mathbf x(t_{k-1}) \ra \mathbf x', \mathbf z(t) | \boldsymbol \alpha ) dt} }  
	\right) \nonumber\\ &&+ \ln p(\boldsymbol \alpha) +C \nonumber \\
	&=&\sum_{k\in K_T} \left( 
	\ln \lambda(\mathbf x(t_{k-1}) \ra \mathbf x(t_k), \mathbf z(t_k) | \boldsymbol \alpha )
	-   \sum_{x'} \int_{t_{k-1}}^{t_k} \lambda(\mathbf x(t_{k-1}) \ra \mathbf x', \mathbf z(t) | \boldsymbol \alpha ) dt  
	\right) \nonumber\\ &&+ \ln p(\boldsymbol \alpha) +C, \label{likelihood}
\end{eqnarray}
where we used Eqs~\eqref{eq_path} and \eqref{eq_exp} and we denoted by $C$ some constant independent of $\boldsymbol \alpha$. An analogous log-likelihood function can be also written for a group of individuals with pairwise interactions. Here, $p(\boldsymbol \alpha)$ is the prior over parameters, which we take to be uniform. 
Intuitively, the first term in Eq~\eqref{likelihood} favors higher rates when more transitions are observed, while the second term favors lower rates when the integrated transition probabilities over intervals become large. The first term depends on the observed transitions, whereas the second term depends on the trajectory between them, accounting for the transitions that could have happened but did not. When dynamical laws are a priori unknown, the sampling of trajectories has to be sufficiently fine to obtain a good quadrature approximation for the integral terms.

\section{Examples of Animal Motion} \label{sec_examples}

\begin{figure}[h]
	\centering
	\includegraphics[width=0.8\textwidth]{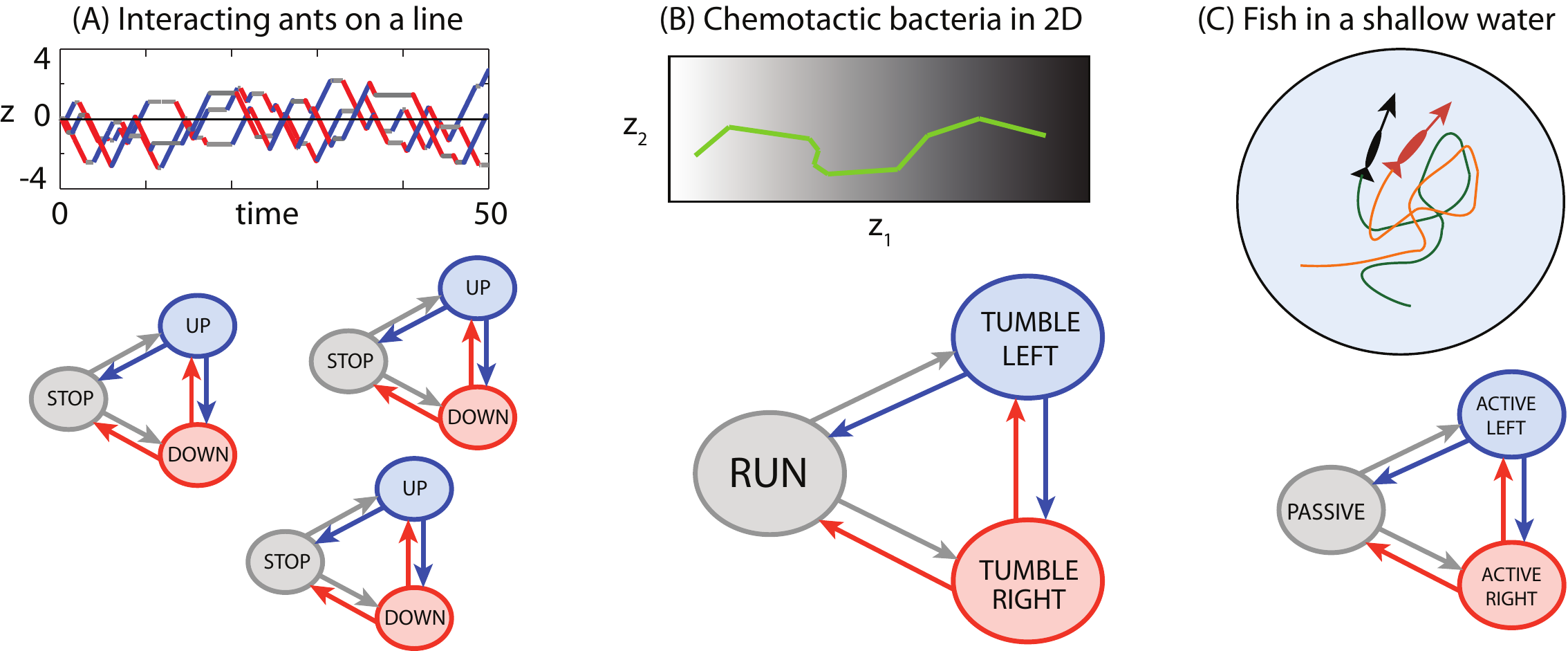}
	\caption{{\bf Examples of animal motion studied in this paper.} The examples vary in complexity, in the dimension of the physical / behavioral space, and in the number of interacting individuals. 
	{\bf (A)} Single ant or several interacting ants with three behavioral states moving in 1D.
	{\bf (B)} Bacteria climbing a chemical gradient in 2D using a run-and-tumble motion. The position space is three dimensional (two spatial coordinates plus the direction). The ``tumble state'' in which bacteria perform directional random walk is a compound state constructed from quick random switches between ``tumble right'' and ``tumble left'' states with deterministic laws of motion.
	{\bf (C)} Two tracked interacting zebrafish moving in 2D in a shallow water tank. The rates of switching between the three kinematic states are assumed to depend on up to three explanatory variables. }
	 \label{fig_examples}
\end{figure}

Here we demonstrate three applications of the inference for simple models of animal behavior. The studied examples, schematized in Figure~\ref{fig_examples}, differ in the number of interacting animals, in the spatial dimension in which the motion unfolds, and in the complexity of  animal motion that we seek to describe. The examples considered are very simplified toy models, whose purpose is not to provide a quantitative match to any particular dataset; rather, we try to showcase the versatility of the modeling framework and the application of our inference and simulation methodology. 

\subsection{Ant motion on a line} \label{ssec_1ant}
In the first example we consider the simplest model of a single ant that is moving along a line. Its motion is represented by three behavioral states, $\mathbf x = \{ x_\text{stop}, x_\text{up}, x_\text{down} \}$, which correspond to motion at a constant velocity in one of the available directions and to a resting state. The equations of motion are
\begin{eqnarray} \label{eq_ODE1ant}
	\frac{dz}{dt} = f(z,x) = 
	\begin{cases} 
		\phantom{-}1\,, & x = x_\text{up}\\
		\phantom{-}0\,, & x = x_\text{stop}\\
		-1\,, & x = x_\text{down}
	\end{cases}\,.
\end{eqnarray}
The changes of the behavioral state are considered random, resulting in a stochastic motion of the ant along the line. The transition rates depend on the position of the ant and have the form
\begin{eqnarray} \label{eq_rates_1ant}
	\lambda(x\ra x',z) = \exp \left( \sum_i \alpha^{(1)}_i(x \ra x') \ph_i(z) \right)\,,
\end{eqnarray}
where each perceived ``stimulus'' $\ph_i(z)$ is modulated by the weight $\alpha_i$. For $\ph_i(z)$ we use the tiling functions in the position space $z\in [-4,4]$, i.e. 
characteristic functions 
of non-overlapping intervals of equal size covering the position space, as in Figure~\ref{fig_tiling}A.

\begin{figure}[]
	\centering
	\includegraphics[width=0.7\textwidth]{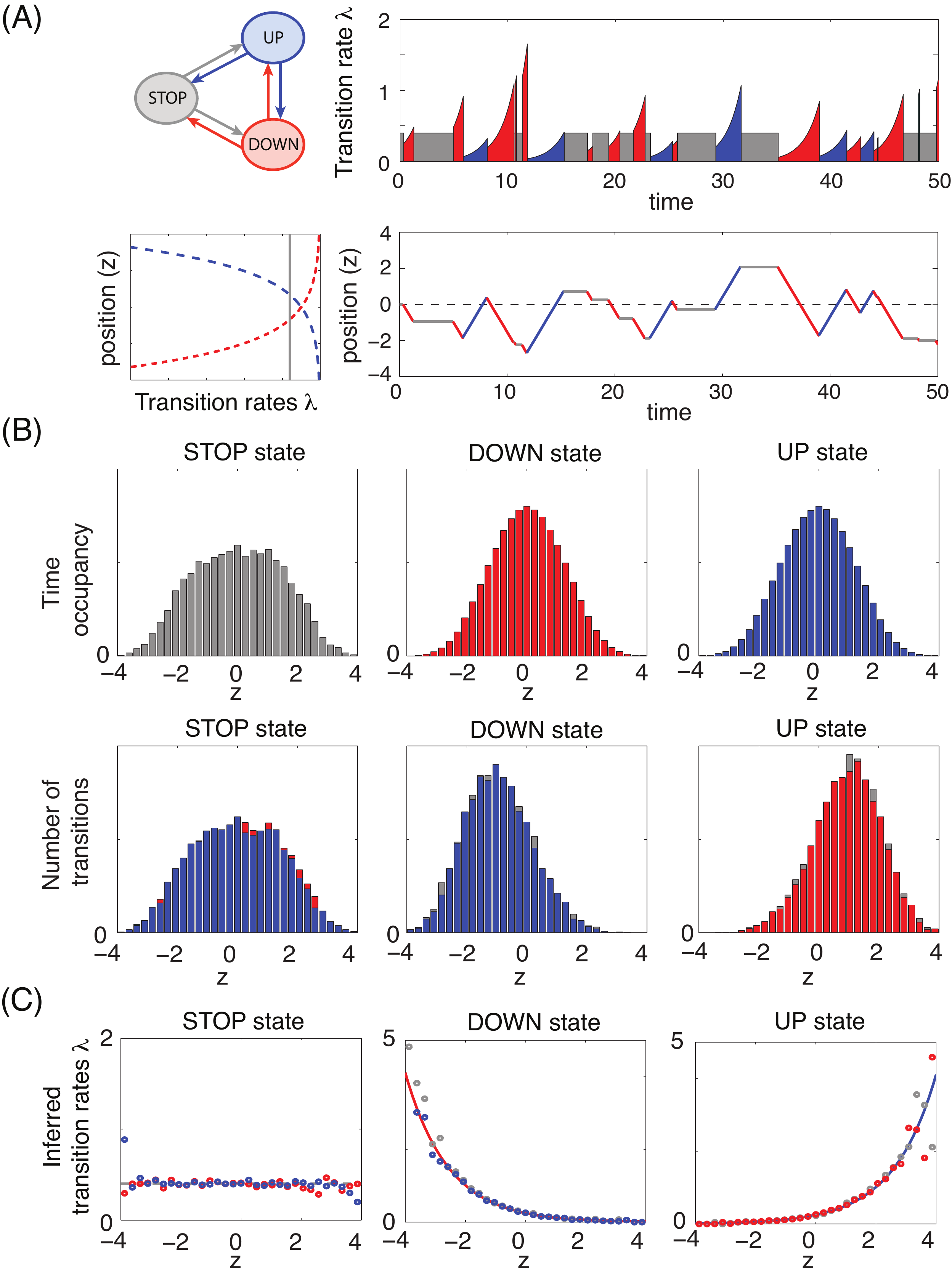}
	\caption{
	{\bf Motion of a single ant on a line.}
	{\bf (A)} Animals switch between three behavioral states: STOP (gray), UP (blue), and DOWN (red).
	The transition rates have a form 
	$\lambda(x \ra x',z) = \exp [\alpha^{(1)}_1(x\ra x') + \alpha^{(1)}_2 (x \ra x') z]$ with 
	$\alpha^{(1)}_1(S\ra x') = \ln(0.4)$, 
	$\alpha^{(1)}_2(S\ra x') = 0$, 
	$\alpha^{(1)}_1(U\ra x') = \alpha^{(1)}_1(D\ra x')= \ln(0.25)$ and 
	$\alpha^{(1)}_2(U\ra x') = -\alpha^{(1)}_2(D\ra x')  = 0.7$. 
		A short segment of a sample stochastic trajectory, following Eq~\eqref{eq_ODE1ant},  is shown 
	together with instantaneous transition rates from the current state.
	 {\bf (B)} Because tiling functions are used to represent position, the sufficient statistics for the inference are conditional histograms.
	 First row shows the histograms of the position given behavioral state during the whole simulation.
	 Second row shows the number of transitions from a given behavioral state in each bin. 
 	 Since from each state transitions can happen into two other states, each of the panels in the second row shows two 
 	 (almost identical) histograms that correspond to two different color-coded target states.
	{\bf (C)}  Inferred transition rates (dots; color denotes the target state) 
	 from the simulated stochastic trajectories 
	 (100 trajectories for $t\in[0,500]$), compared with the exact transition rates (solid curves; color denotes the current state). 
	 Since each state can transition into two other states, the inference provides separate estimates for each transition, 
	 shown as two sets of dots of different color. The true rates for two target states are equal. 
		}
	\label{fig_1ant}
\end{figure}

A long stochastic trajectory of the ant motion following Eqs~\eqref{eq_ODE1ant}-\eqref{eq_rates_1ant}  was simulated with chosen transition rates that depended on the position of the ant; a fraction of the trajectory and the transition rate details are shown in Figure~\ref{fig_1ant}A. This trajectory was then used to infer the transition rates, that were considered unknown in the inference step. 
Due to our choice of the tiling functions $\{ \varphi_i \}$, the inference decouples into independent likelihood optimizations for each bin. Crucially, the inference does not require knowledge of the detailed trajectories, just of their sufficient statistics, i.e., the distribution of ant positions at the transition times and the overall time occupancy of every bin, shown in Fig~\ref{fig_1ant}B. This is because the log-likelihood function can be also written as
\begin{eqnarray}
	L(\boldsymbol \alpha) &=&
	\sum_{i} \sum_{x,x'}
	\alpha^{(1)}_{i}(\mathbf x \ra \mathbf x', \mathbf z_{i}) 
	\mathcal{S}_1(\mathbf x \ra \mathbf x', \mathbf z_{i}) \nonumber\\
	&+& \sum_{i} \sum_{x,x'}
	\exp \left[ \alpha^{(1)}_{i}(\mathbf x \ra \mathbf x', \mathbf z_{i}) 
	\right] \mathcal{S}_2(\mathbf x, \mathbf z_{i}) 
	+ \ln p(\boldsymbol \alpha) +C, \label{likelihood_hist}
\end{eqnarray}
where $\mathcal{S}_1$ and $\mathcal{S}_2$ represent the total number of transitions $x \ra x'$ at each bin and the total time spent in each bin, respectively. Here $\mathbf z_i$ denote the center of the $i$-th tiling function, i.e. an average position when in the $i$-th bin. 

The inferred rates, shown in Figure~\ref{fig_1ant}C, closely captured the true rates when the space was well sampled. For our choice of transition rates the sampling coverage decreases exponentially as $|z|$ increases, so rate estimates at higher values for $|z|$ are less reliable; nevertheless, low sample numbers in one bin do not affect the accuracy of the estimate in other bins. 
Generally, the quality of inference is limited by the available data. The ability to sample many long motion trajectories clearly leads to better parameter estimates. A less obvious factor that also influences the results is the sampling frequency, which sets the limit to the temporal precision with which behavioral transitions can be estimated, and to the quality of the approximation for the integral term in Eq \eqref{likelihood}. We explore the consequences of undersampling in Appendix~\ref{ssec_sampling}.

\begin{figure}[]
	\centering
	\includegraphics[width=0.7\textwidth]{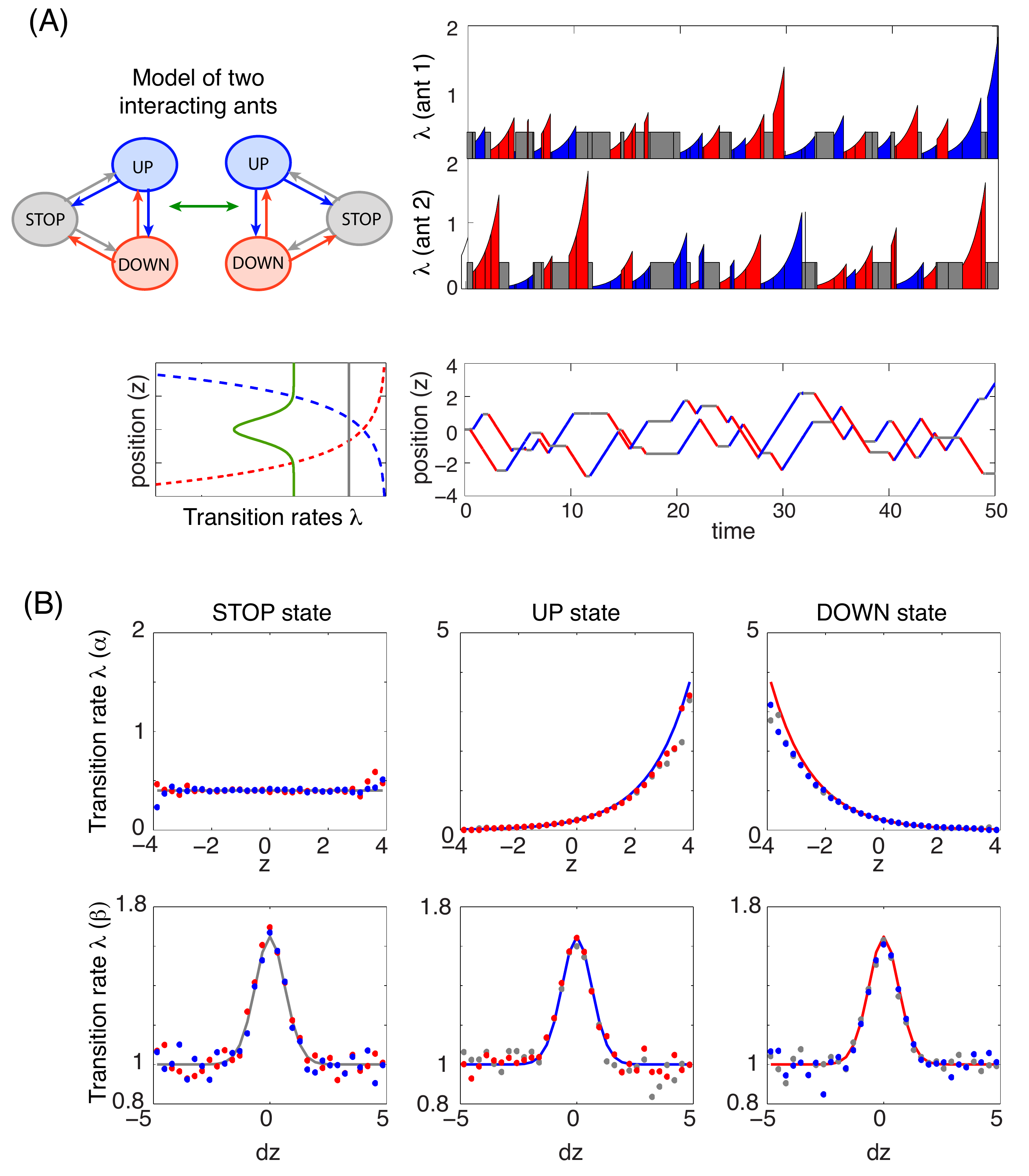}
	\caption{
	{\bf Two interacting ants.} The ants follow dynamics of Eq~\eqref{eq_ODE1ant} where 
	the stochastic switching between the states follows transition rates
	$\lambda(x \ra x', \mathbf z) = 
	\exp[{\alpha^{(1)}_1(x\ra x') + \alpha^{(1)}_2 (x \ra x') z + \sum_{n\neq n'} \alpha^{(2)}_1(x \ra x')  \psi(z_{n'}-z_n)}]$
	with $\alpha^{(1)}_1$ and $\alpha^{(1)}_2$ the same as for the single ant in Fig~\ref{fig_1ant} and the interaction strength
	$\alpha^{(2)}_1(x\ra x') = 0.5$. 
	The interaction with a kernel $\psi(d) = e^{-d^2}$ modulates the rates multiplicatively by a factor which increases all transition rates
	when ants are close to each other and leaves them unchanged when they are sufficiently distant.
	{\bf (A)} Short stochastic simulation of two interacting ants, showing also the instantaneous transition rates from the current state.
	The interaction modulating factor  $\exp(0.5\psi(d))$ is shown together with other transition rates (green) using the 
	same axis.
	{\bf (B)} The first and the second rows show the inferred transition rates (dots) 
	for the parameters $\alpha^{(1)}$ and $\alpha^{(2)}$, respectively, compared with the exact rates (solid curves). 
	The inference is based on 500 simulated trajectories, each for $T\in [0,500]$. 
	We used the tiling functions in the $z$ and $\Delta z = z_n - z_{n'}$ space and 
	a penalization function of the form $\sum_j(\exp(\alpha^{(2)}_j)-1)^2$ to enforce vanishing interaction 
	outside of the range $|\Delta z|>2$. Penalization term also avoids a degeneracy of the rates, ensuring existence of a 
	unique solution of the inference problem. 
	}
	\label{fig_2ants}
\end{figure}

The single ant model can be extended to multiple interacting ants. In this case, the transition rates of the $n$-th ant depend not only on its own position, but also on the distance between the ants, as illustrated in Fig~\ref{fig_2ants}A.
To model the interaction between a pair of ants, we introduced an interaction kernel, $\psi(d) = e^{-d^2}$, that depended on the mutual distance between the two ants, $d=|\Delta z|$. Smaller distances lead to a larger additive contribution to the argument of the nonlinear function $g(\cdot)$ in Eq~\eqref{eq_rates}, and since $g$ in our model is an exponential function, the interaction modulated all other transition rates by a multiplicative factor. In effect, when two ants were close by, all their transition rates increased, leading to fast switching among the behavioral states.

To perform inference in the case of two interacting ants, we optimized the likelihood written out in terms of the summary statistics, with the results shown in Fig~\ref{fig_2ants}B. Summary statistics in this case are the joint distribution of the time spent at each position $z$ and at separation $\Delta z = z_{n'}-z_n$ between the $n$-the and the $n'$-th ant, and the marginal distributions of the transition counts as functions of position $z$ and distance $\Delta z$. In case of many interacting ants, the summary statistics may become too high-dimensional, prompting us to revert to  Eq~\eqref{likelihood}, where the likelihood is evaluated directly over the trajectories.

\subsection{Bacterial chemotaxis} \label{ssec_bacteria}
Many bacteria have the ability to climb gradients of chemoattractant chemicals, for instance, when searching for food. One strategy for climbing such gradients is the so-called run-and-tumble motion~\cite{berg1972}. Here, periods of motion in a straight line with nearly constant velocity, called ``runs,'' are interspersed with ``tumbles,'' events when a bacterium randomly reorients itself. The net motion bias towards the gradient source ensues because the rate at which tumbles are initiated depends on whether the bacterium has recently been moving along, or against, the gradient. 

The run-and-tumble paradigm does not map directly into our framework, because the dynamics in the tumble state is stochastic: it can be approximated as a directional diffusion or a random walk in orientation. We can, however, propose a three-state model that is able to capture the stochastic tumble state, as we show in Appendix~\ref{ssec_CG}. The three behavioral states that we introduce for this purpose are: (i) a run state, where the bacteria move at a constant speed and direction; (ii+iii) left/right tumble states, where the bacteria are stationary but rotate to the left or right at a constant angular velocity. The bacterial motion is described by three variables: position $z_1=x$, $z_2=y$, and angular direction $\varphi$ in the 2D plane. Kinematic variables  obey the following dynamical equations:
\begin{eqnarray}
	&\text{For }& x = x_\text{run}: \quad z_1' = v\cos \alpha\,,  \quad z_2' = v\sin\alpha\,, \quad \ph' = 0\,, \label{eq_ODE_RT1}\\
	&\text{For }& x = x_\text{left/right}: \quad z_1' = 0\,, \quad z_2' = 0\,, \quad \ph' = \pm \omega \label{eq_ODE_RT2}\,,
\end{eqnarray}
where $v$ is the speed of the bacterium in the run state and $\omega$ is the angular velocity in the tumble state, both assumed constant. Since the transitions between the states are stochastic, the length of the trajectory in the run state as well as the net rotation angle in the tumble phase are random variables. The key idea behind the three state model with $\mathbf{x} = \{ x_{\rm run}, x_{\rm left},x_{\rm right}\}$ is that if the switching rates between the ``tumble left'' ($x_{\rm left}$) and ``tumble right'' ($x_{\rm right}$) states are fast, the rapid succession of random switches between the two deterministic left/right tumble states will generate a random walk in orientation, effectively simulating a single, stocahstic, compound ``tumble'' state, equivalent to the one described in chemotaxis.

\begin{figure}[]
	\centering
	\includegraphics[width=\textwidth]{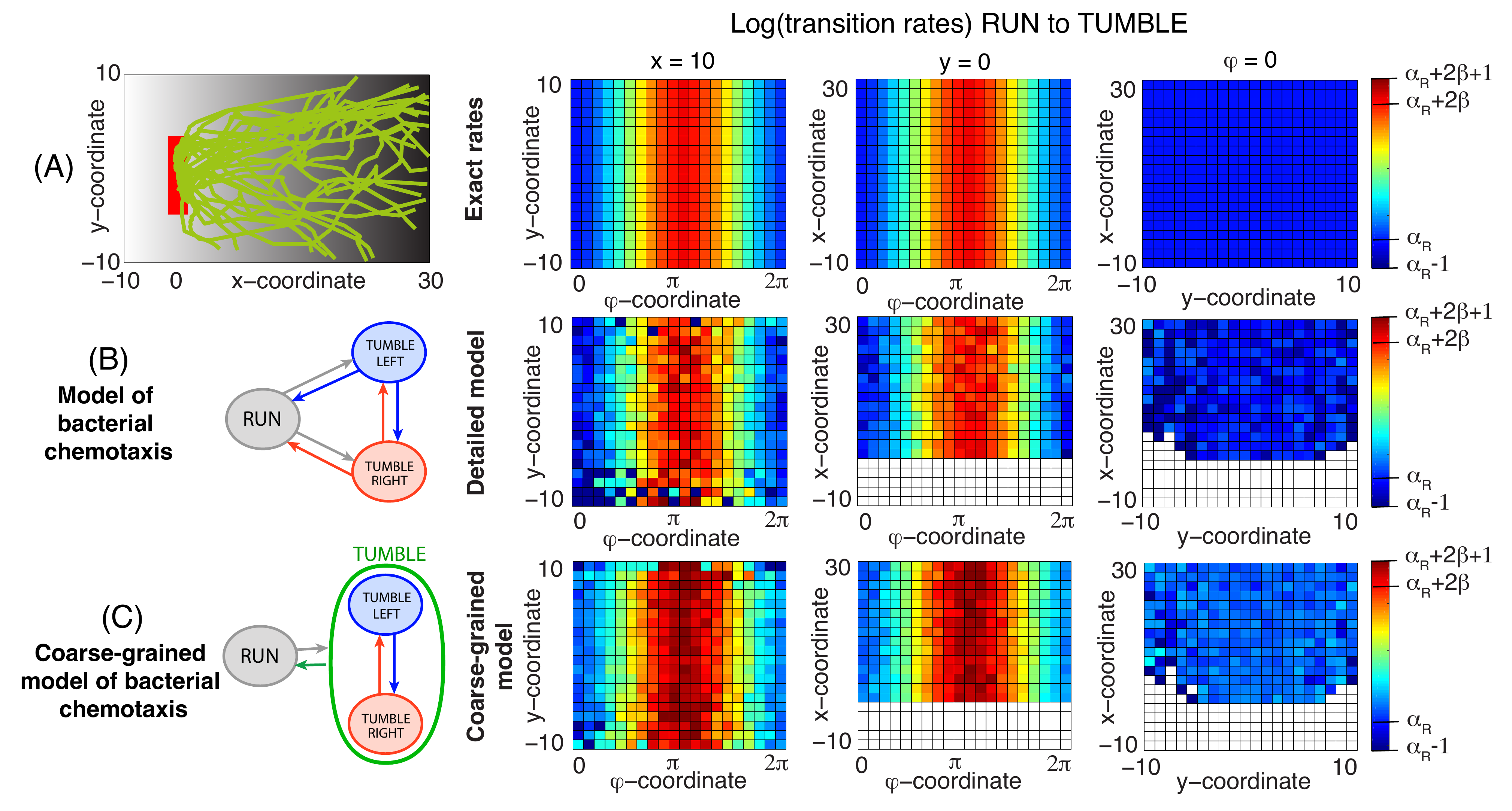}
	\caption{
		{\bf Bacterial chemotaxis.} Bacteria execute run-and-tumble motion to climb the chemoattractant gradient 
		(here, in the direction $\ph^*=0$), starting at a random position within the red region. 
		We assume the following transition rates:
		$\lambda(x_\text{run}\ra x_\text{left/right})= \exp(\alpha_R + \beta f(\ph) )$,
		$\lambda(x_\text{left/right}\ra x_\text{run})= \exp(\alpha_{L} )$,
		$\lambda(x_\text{left/right}\ra x_\text{right/left})= \exp(\alpha_{LR} )$,
		where $\alpha_R = \alpha_L = \alpha_{LR} = \log(0.3)$, $\beta = 3$,  $v = 2$,  $\omega = 1$. The constants
		$\beta$, $v$, and $\omega$ represent the strength of the chemical gradient, the speed of the bacterium in the run phase, 
		and the angular velocity of the bacterium in the tumble state, respectively.
		The transition rate from the run state depends on the internal angle of the bacteria through a response 
		function $f(\ph)$. For simplicity we take $f(\ph) = 1+\cos(\ph-\ph^\ast-\pi)$, which is maximal for an angle 
		antiparallalel to $\ph^\ast$, where $\ph^\ast$ determines unobserved location of the source of 
		chemoattractant.
		{\bf (A)} Individual trajectories of a run-and-tumble bacterial motion in a chemical gradient simulated by a 
		random switching between three deterministic behavioral states (run, tumble left, tumble right), following the 
		dynamics of Eqs~\eqref{eq_ODE_RT1}-\eqref{eq_ODE_RT2}. At right, exact transition log-rates are shown for three 
		different cross-sections in the position space: $(\ph,10,y)$, $(\ph,x,0)$ and $(0,x,y)$. 
		{\bf (B)} Transition (logarithmic) rates inferred from the simulated  trajectories 
		(2000 trajectories that start at location $x=0$) assuming a model with three deterministic states.  
		{\bf (C)} Inferred transition (logarithmic) rates for the coarse-grained model of bacterial chemotaxis with one stochastic, 
		compound ``tumble'' state.
		Regions poorly sampled by simulated trajectories are shown in white. The colorscale ranges between 
		$[\alpha_R-1,\alpha_R+2\beta+1]$, where $\alpha_R$ and $\alpha_R+2\beta$ are the minimum and the maximum of the true
		log-rates.
	}
	\label{fig_RT}
\end{figure}

We consider an example where the transition rates from the run state into the tumble states, $\lambda = \lambda(\varphi,x,y)$, depend on the position of the bacterium and its orientation, as displayed in Fig~\ref{fig_RT}A: the rate of exiting the run state is low when the bacterium is oriented parallel to the chemoattractant gradient and high when the bacterium is antiparallel. Simulated trajectories, shown in Fig~\ref{fig_RT}A, show that this bias gives rise to motion in the gradient direction. 

To perform inference, we discretized kinematic variables, $\mathbf z$, into $N_\ph \times N_x \times N_y$ bins, and inferred the transition rates bin-by-bin from simulated stochastic trajectories. This approach made no a priori assumptions about how the rates varied with the coordinates. 
As shown in representative cross-sections in Fig~\ref{fig_RT}B, the inference led to a good estimate of the rates in the parameter regime which was sampled sufficiently by the 2000 simulated trajectories. In a real-world application, further assumptions about the smoothness of the rate dependence, either by a proper choice of position representations or by explicit smoothness regularization, and adaptive discretization selected by, e.g., cross-validation, could further boost the performance of inference given limited data.

Real observations of bacterial chemotaxis do not differentiate between the deterministic ``tumble left'' and ``tumble right'' states: in the limit of fast transitions between the two tumble states (relative to the transitions into the run state), the two tumble states merge into a single stochastic ``tumble'' state, in which the bacterium performs directional diffusion. We therefore asked if we can coarse-grain our framework analytically to infer the dependency of transition rates between the run state and this new, compound ``tumble'' state, directly. In Appendix~\ref{ssec_CG} we show that this is indeed possible under moderate assumptions. Direct inference of the two-state coarse-grained model (with a deterministic ``run'' and a stochastic ``tumble'' state) tends to give superior performance compared to the three-state model (with deterministic ``run,'' ``tumble left,'' and ``tumble right'' states), as shown in Fig~\ref{fig_RT}C, because the coarse-graining acts as an implicit regularization that integrates over some of the parameters of the three-state model. On the other hand, the coarse-grained model is applicable only under the time-scale separation and so microscopic details about behavioral transitions between ``tumble left'' and ``tumble right'' states, as well as the information about angular velocity $\omega$ in these states, are lost or are poorly constrained by the data. Using coarse-graining to define new compound behavioral states with stochastic laws of motion, and performing inference on the corresponding coarse-grained models directly, as demonstrated in the chemotaxis example, should significantly expand the modeling domain of our framework.

\subsection{Motion of fish} \label{sec_fish}

In the last example we show how to proceed from raw data, where the true (generating) model is unknown, to a set of infered models that can be compared using model selection techniques. We do not aim to construct a perfect model of the animal motion; rather, our goal is a practical demonstration of the applicability of our approach to real data. The data was collected by \cite{Harpaz2017} where tracked zebrafish are swimming in a shallow water tank of a circular shape, roughly a meter in diameter. The duration of the recordings, performed in two replicates, was about 30 minutes and the sampling rate was 100 Hz. We will see that this provides enough data to extract the basic features of fish motion.

\subsubsection{Identification of states and laws of motion} \label{ssec_Fmodel}

\begin{figure}[]
	\centering
	\includegraphics[width=0.65\textwidth]{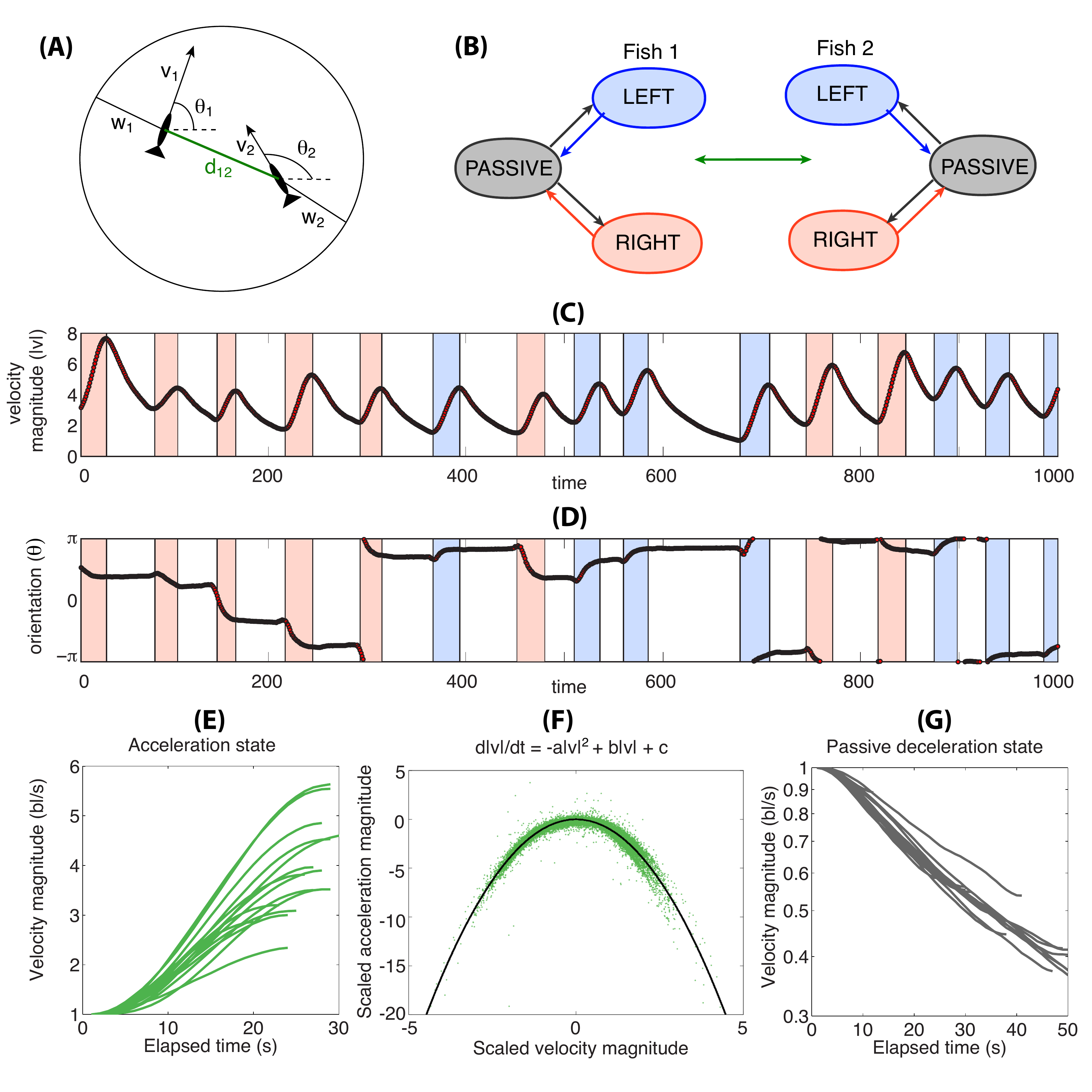}
	\caption{{\bf Motion of two fish in a circular shallow water tank. }
		{\bf (A)} Each fish is characterized by a position $(x,y)$, velocity $\mathbf v$, 
		orientation $\theta$ and acceleration $a$.	
		{\bf (B)} Diagram of behavioral state transitions for the two fish with interaction.
		{\bf (C-D)} A time window of length 1000 s shows the velocity and orientation of one fish (second fish not shown) from the
		data in~\cite{Harpaz2017}. The trajectory shows alternating regions of acceleration 
		(blue = turning left, red = turning right) or deceleration (no shade), consistent
		with the states marked in B.
 		{\bf (E)} Velocity traces in the accelerating phase (data from C), shifted to the same initial value, 
 		have a sigmoidal functional form. Acceleration can be fitted empirically to a 
 		quadratic function, $d|v|/dt = -a|v|^2 + b|v|+c$ where $|v|=\sqrt{(\dot x)^2+(\dot y)^2}$. 
 		This is shown in {\bf (F)} for data containing 3000 accelerating intervals. 
		Data from each accelerating window (dotted green) is fitted separately 
		and then shifted and rescaled to a normal form $|a|=-|v|^2$ (black). 
		{\bf (G)} Passive state shows an exponential decay of the velocity due to friction, 
		evident by rescaling decelerating trajectories to the same initial value and plotting them on a log-linear scale. 
		}
	\label{fig_fishmain}
\end{figure}

Basic physical laws dictate the natural minimal set of kinematic variables, i.e., the position, velocity and acceleration of each animal. The motion of a fish in the shallow water tank is essentially two-dimensional, with position $\mathbf z_i = (x_i,y_i)$ where $i=1,2$ is the index of the fish (with analogous notation used for the velocity, $\mathbf{v}_i$, and acceleration, $\mathbf{a}_i$), so that:
\begin{eqnarray}
	\frac{d\mathbf z}{dt} = \mathbf v\,,\quad \frac{d\mathbf v}{dt} = \mathbf a\,.
\end{eqnarray}
Figure~\ref{fig_fishmain} shows a typical segment of the trajectory and, in particular, highlights the repeated fluctuations in the velocity magnitude, consistent with the findings of \cite{Harpaz2017}.  
These fluctuations are generated by a stereotypical force, caused by the tail beats and fin flips of the fish. Notice also that each acceleration event is accompanied by a dynamic adjustment of the angle $\theta_i$, which measures the orientation of the fish, while in the deceleration phase the orientation stays unchanged. The data thus indicate that individual fish can be characterized by distinct behavioral states that correspond to the increase and decrease in the velocity of the fish. We call these states either active, where the motion of the fish results in a positive acceleration and in the change of orientation, or passive, where the dynamics are caused by the water friction without any active contribution of the fish. We identify the following laws of motion for the two kinds of states that hold to a good approximation:
\begin{itemize}
	\item {\it Passive deceleration:} 
		Figure~\ref{fig_fishmain}G indicates an exponential decay of the velocity magnitude in the 
		passive state. Such dynamics are consistent with a passive motion where the fish exerts no force 
		and decelerates solely due to friction.  
		This can be captured by a dynamical law for the magnitude of the velocity
		\begin{eqnarray}
			\frac{d|v|}{dt} = -\kappa |v|\,. \label{eq_ODEv_passive}
		\end{eqnarray} 
		with an exponentially decaying solution with a rate $\kappa$, representing the friction coefficient.
		The orientation of the fish in the pasive state satisfies 
		$\frac{d\theta}{dt} = 0$. 
	\item {\it Active acceleration, left/right turn:} The fish exerts a pulse-like force as a result of its motion, as in
		Fig~\ref{fig_fishmain}E. This motion can be accurately fitted by the following law:
		\begin{eqnarray}
			\frac{d|v|}{dt} = -k_2|v|^2+k_1|v| + k_0\,, \label{eq_ODEv_active}
		\end{eqnarray}
		as indicated by Fig~\ref{fig_fishmain}E-F. Here, an initial exponential increase in speed
		due to the force generated by the fish body motion is followed by saturation due to frictional dissipation of energy.  
		The orientation of the fish changes in the active state,
		with an approximately exponential relaxation to a target $\theta^*$ from the initial value 
		$\theta(0) = \theta_0$. The dynamics can be approximated by a linear mean-reverting process
		\begin{eqnarray}
			\frac{d\theta}{dt} = \tau(\theta^\ast-\theta)\,. \label{eq_ODEtheta_active}
		\end{eqnarray}
\end{itemize}
While the dynamics obviously
differ between the passive and active states, they formally differ also within the active state(s). This is because the dynamics in the active states are only fully specified when the five event-specific parameters, i.e. $\theta^\ast-\theta_0$, $\tau$, $k_0$, $k_1$, and $k_2$, are known. The data suggest that these parameters have relatively narrow distributions. While in principle the states should be parametrized by these five parameters, yielding a large total number of total behavioral states, we disregard this dependence here and use only three behavioral states: passive and active left/right, as shown in Figure~\ref{fig_fishmain}B. The dynamics in Eqs
\eqref{eq_ODEv_active} and \eqref{eq_ODEtheta_active} would then be approximated by typical values of the parameters ($\langle \theta^\ast-\theta_0 \rangle$, $\langle \tau \rangle$, $\langle k_0 \rangle$, $\langle k_1 \rangle$ and $\langle k_2 \rangle$) or alternatively, these values can be generated stochastically from a suitable distribution. The data further suggest that the behavioral states alternate and that two active states are separated by a passive state.

\subsubsection{Inference of fish behavior} \label{ssec_Finference}
\begin{figure}[]
	\centering
	\includegraphics[width=0.85\textwidth]{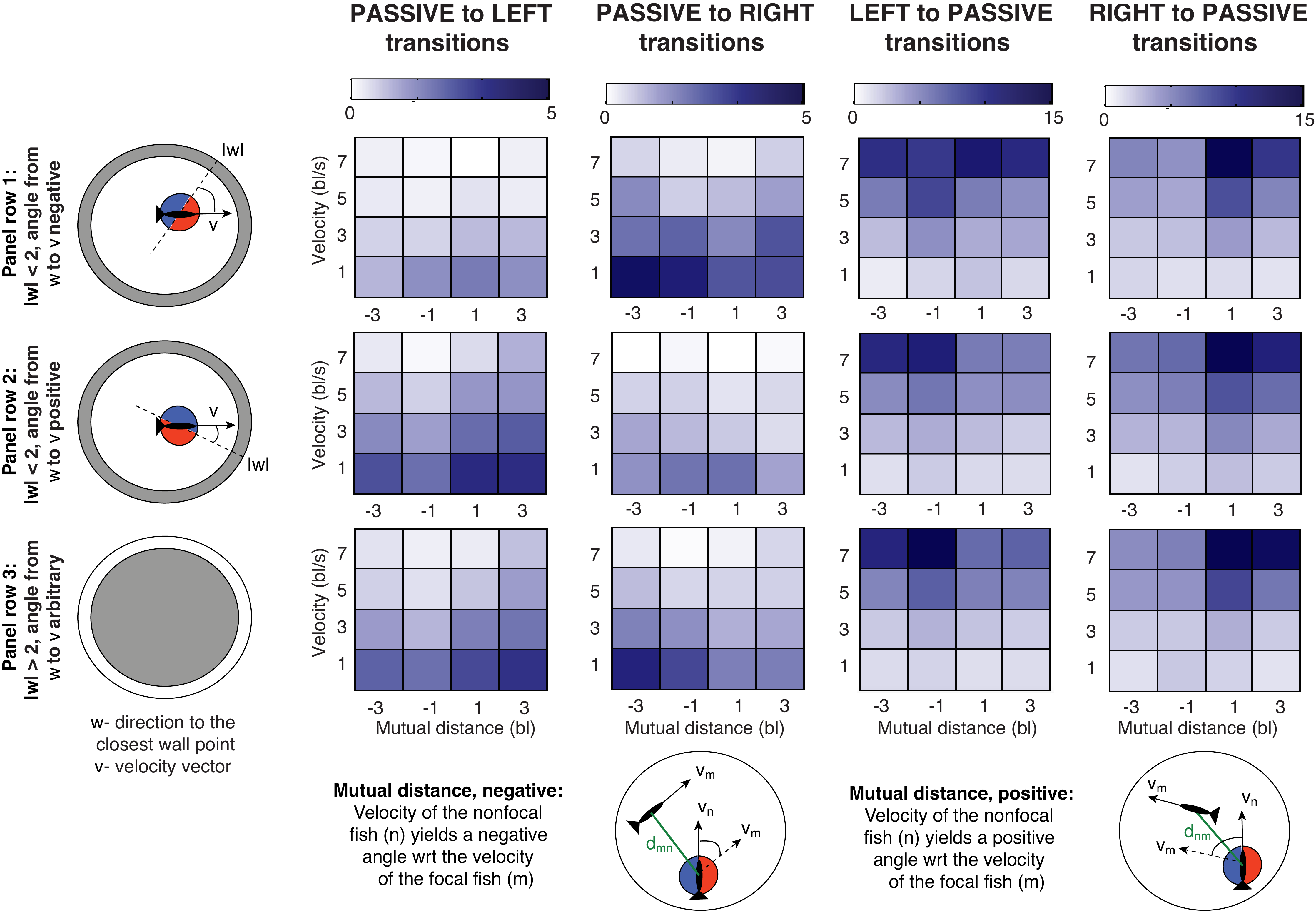}
	\caption{
	{\bf Transition rates inferred from tracked zebrafish data for the model with three explanatory variables.}
	Transition types are indicated in column headings. For each transition type,  
	$\lambda(s\ra s')$, from state $s$ to $s'$, the rates are shown by color intensity (colorbar on top), as a function of
	mutual distance $u^3$ (horizontal axis), velocity magnitude $u^1$ (vertical axis) 
	and wall distance $u^2$ (three rows, see legends at left). 
	We used bin centers $\{1,3,5,7\}$ for the velocity magnitude $u^1$, bin centers $\{-3,-1,1,3\}$ 
	for the mutual distance $u^3$ and the wall distance is split to three bins: 
	$ u^2 \in [-2,0)\,, \, [0,2]$, and $|  u^2 |>2$. 
	The rates were inferred jointly from two experiments, each with two fish; the rates were assumed to be the same for all the fish.
	}
	\label{fig_Finfer}
\end{figure}

After identifying the three behavioral states and their laws of motion we now turn our attention to the factors that modulate behavioral state transitions. The first goal is to identify the suitable explanatory variables $\mathbf u$ for the transition rates. These should account for the behavior of the focal fish, the influence of the environment, but also the interaction between the fish.  The fish spend a significant amount of time in a close proximity of the tank walls but also that fish often swim short distance apart from each other. To reflect this we select the following explanatory variables:
\begin{itemize}
	\item 
	{\it Velocity magnitude: $u^1(\mathbf z_n) = |\mathbf v_n| = \sqrt{(\dot x_n)^2+(\dot y_n)^2}$}, where the velocity 
	of the $n$-th fish $\mathbf v_n = (\dot x_n,\dot y_n)$ is approximated from the data. 

	\item 
	{\it Wall distance with a sign: $u^2(\mathbf z_n)  
	= \sigma|\mathbf w_n|$.} 
	The quantity we use to capture the wall distance is derived from a vector  $\mathbf w_n$, 
	leading from the focal fish position to the closest point at the fish tank wall. 
	The wall distance $|\mathbf w_n|$ provides information on the proximity of the wall, independent of the orientation of the fish. 
	To encode also the direction of swimming along the wall we multiply $|\mathbf w_n|$ by a sign $\sigma$, 
	which is positive/negative if the fish swims anticlockwise/clockwise along the wall. 
	The sign can be expressed as a vector product of the wall direction with the direction of swimming, 
	i.e., $\sigma = \text{sgn}(\mathbf w_n \times \mathbf v_n)$.
	
	\item 
	{\it Mutual alignment: $u^3(\mathbf z_n,\mathbf z_m)  = 
	\rho d_{nm}$.}
	The interaction between the fish $n$ and $m$ depends on their mutual distance and their alignment. 
	We define $d_{nm} =  |\mathbf z_n- \mathbf z_m|$ 	
	as a distance between the fish, modulated by a sign $\rho$. The distance (from the perspective of the $n$-th fish) is positive 
	if the directed angle from $\mathbf v_n$ to $\mathbf v_m$ is in $[0,\pi]$, zero, if the fish are perfectly aligned, 
	and negative otherwise, i.e., $\rho = \text{sgn}
	(\mathbf v_n \times \mathbf v_m)$. Quantity $\rho d_{nm}$, measuring alignment between the orientation of the fish, 
	is an asymmetric quantity, i.e., changing the reference fish changes the sign but not the magnitude of $d_{nm}$, see Fig~\ref{fig_Finfer}. 
\end{itemize}

Next we express the transition rates of the $n$-th fish similarly as in Eq~\eqref{eq_rates}:
\begin{eqnarray}
	\lambda(s \ra s',\mathbf z) = \exp \left( \sum_
		{\substack{m=1\\
                  m\neq n }}
                  ^N \sum_{i}^{N_{u^1}} \sum_{j}^{N_{u^2}} \sum_{k}^{N_{u^3}} \alpha_{ijk}(s\ra s') \varphi_{ijk}(\mathbf u) \right). \label{eq_Frates}
\end{eqnarray}
where $\mathbf u(\mathbf z_n,\mathbf z_m) = (u^1,u^2,u^3)$ are the specified key kinematic variables and $N_{u^1}$, $N_{u^2}$, $N_{u^3}$ the number of discretization bins for each. 
This approach does not assume any particular decomposition of the rates, unlike Eq~\eqref{eq_rates} where the no-interaction terms with $\alpha^{(1)}$ and interaction terms with $\alpha^{(2)}$ are assumed to sum additively in the exponent of the transition rates. Thus this ansatz is more general than in the synthetic example of ant motion. Indices $n$, $m$ in Eq~\eqref{eq_Frates} indicate the identity of the focal and non-focal fish, respectively, and the summation goes through all perceived stimuli, captured by the kinematic variables $(u^1,u^2,u^3)$, which are functions of position of the focal and non-focal fish.  The basis functions are chosen to be the tiling functions $\varphi$ in the joint 3-dimensional space of $u^1$ and $u^2$ for single-fish position coordinates, and also $u^3$ for the pairwise-interaction term.

\begin{figure}[h]
	\centering
	\includegraphics[width=0.7\textwidth]{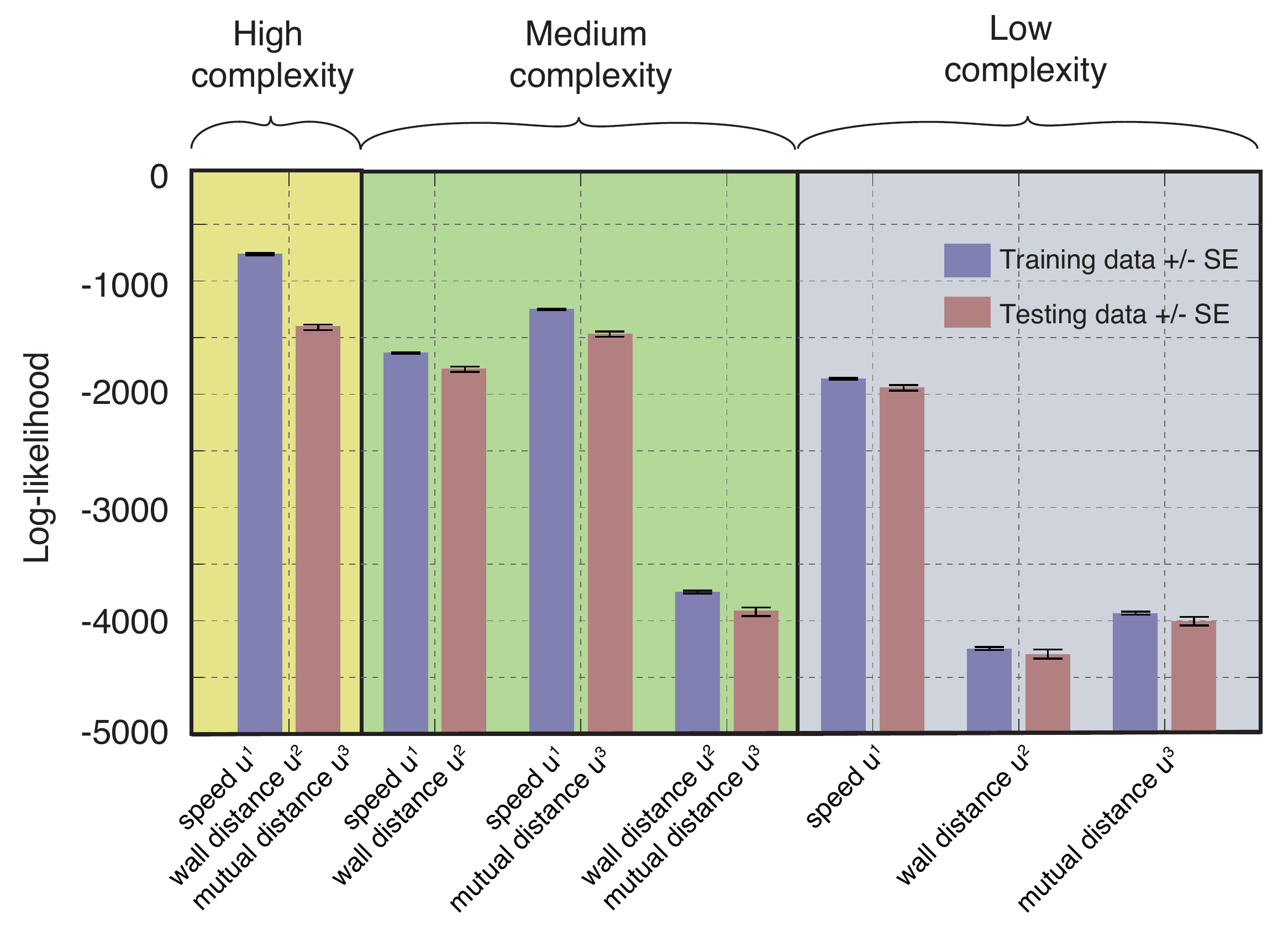}
	\caption{{\bf Model comparison.}
	Seven models (sorted into the ``high complexity'' model M7 with 3 variables, three ``medium complexity'' models (M4-M6) 
	with 2 variables each, and 3 ``low complexity'' models (M1-M3) with one variable each) are compared in terms of 
	their log-likelihood on training (blue) and testing (red) data.
	The data used in Fig~\ref{fig_Finfer} was split into segments of 10 s, 
	containing on average 35 state transitions, for a  total  of  180 segments. These segments were then 
	randomly assigned to a training and testing set, with probabilities 0.75 (training set) and 0.25 (testing set). 
	The transition rates were inferred 
	from the training set and tested on the testing set. We generated in total 200 random assignments. 
	Bars are averages over sample sets $\pm$ standard error.
	}
	\label{fig_validation}
\end{figure}

In summary, we consider three explanatory variables $u^1(\mathbf z_n)$, $u^2(\mathbf z_n)$ and $u^3(\mathbf z_n,\mathbf z_m)$. To evaluate the importance of these variables systematically, we construct a hierarchy of seven models by picking different combinations of independent variables (dropping the dependence on $\mathbf z_n$, $\mathbf z_m$ in the notation):
\begin{itemize}
	\item [M1-M3.]  
	Marginal models with only one explanatory variable, i.e., 
	$[\, u^1\,]$ OR $[\, u^2 \,]$ OR $[\,u^3 \,]$.
	\item [M4-M6.]  
	Models with a pair of explanatory variables, i.e., $[\,u^1, u^2\,]$ OR $[\,u^1,u^3\,]$ OR $[\,u^2,u^3\,]$.
	\item [M7.] 
	Model with all three explanatory variables, i.e., $[\, u^1,u^2,u^3\,]$. 
\end{itemize}
The inferred rates for model M7 are shown in Fig~\ref{fig_Finfer}. The systematic dependence of the rates on explanatory variables suggests the following interpretable behaviors of the fish, qualitatively consistent with previous reports in~\cite{Harpaz2017}: 
\begin{itemize}
	\item 
	{\bf Speed dependence.} A passive fish is more likely to transition to an active state 
	when its speed is low, while the active fish is more likely to transition to a passive state when its speed is high. 
	The first statement may be due to a correlation between the speed and the amount of time 
	spent in the passive state, since the fish decelerate in time.
	
	\item 
	{\bf Collision avoidance.} The last two columns of Fig~\ref{fig_Finfer} suggest that a fast moving active fish
	has a tendency to exit the current active turning state when the second fish travels to the same direction. 
	This may be interpreted as a mechanism for avoiding collision between the fish since it applies to large speeds only,
	and does not depend on the distance from the wall.
	
	\item 
	{\bf Wall asymmetry.} A passive fish within a few body lengths of the wall tends to turn in a direction
	to avoid a collision with the wall. This is evident from the contrast in the intensity of the rates between the
	first two panels in the top row and in the middle row in Fig~\ref{fig_Finfer}. 
	The fish swimming to the right/left of the wall will tend 
	to turn to the right/left. Once the fish is far away from the wall (bottom row) this effect vanishes. We observe no 
	similar effect for active fish.
	
	\item 
	{\bf Fish alignment.} There is an asymmetry in the transition rates from the passive to the active state 
	due to alignment of the fish. We observe a large transition rate passive$\ra$right when $u^3<0$ and,
	similarly, a large transition rate passive$\ra$left when $u^3>0$. This implies that the passive fish gets recruited
	by the neighboring fish to become active and align with its neighbor.
\end{itemize}
Our probabilistic models M1-M7 are amenable to systematic model comparison, shown in Fig~\ref{fig_validation}.
While it is evident that the variable $u^1$ individually has 
the largest explanatory power, the inclusion of other variables  also leads to better generalization performance. Given our data, the best model is the complex, three-variable model M7 that includes  self-interaction, effects of environment and the interaction between the fish.

\section{Conclusion}
We introduced a novel approach for understanding the behavior of individual animals or groups of interacting animals. The probabilistic model of animal behavior combines deterministic dynamics, which describe the motion of animals at each of the possible behavioral states, and stochastic switching between these states. Our models are very flexible in terms of what behaviors are being tracked (which is problem specific), in terms of what motions the animals execute (which can be arbitrary deterministic or even stochastic motions), and in terms of which variables influence behavioral state transitions (which we seek to infer). Despite this flexibility, illustrated here by several synthetic and real examples, the forward problem (simulation) and the inverse problem (inference) remain tractable, primarily because the models inherit all the favorable properties of the Generalized Linear Models framework. 
Our framework therefore opens up the possibility to carry out theoretical explorations of, e.g., collective behavior, using forward simulation within the same model class in which inference from data is possible, bringing together two approaches that have in the past interacted rather sparsely. 
Because our models are probabilistic, we can use rigorous tools not only to perform inference, but also to select between classes of models of different complexity and
to identify biologically relevant variables that modulate animal behavior. It is this ability, in particular, that should prove attractive for biological applications.

The practical use of our method consist of several steps, some of which we described in detail. First, when the laws of motion are unknown, as is usually the case, the motion of the animals needs to be recorded at a fine enough sampling rate. Second, distinct behavioral states of interest must be identified on every tracked trajectory. Third, the list of candidate explanatory variables, which the transition rates between the behavioral states may depend on, have to be selected. Fourth, these rates are decomposed into a linear combination of coefficients, to be inferred, and position basis functions, following which the rates are found by solving a convex maximum likelihood optimization problem. Fifth, model selection is performed to find the set of most relevant explanatory variables and, possibly, also the best choice of basis functions. Finally, we can look for a biological interpretation of the switching rates, or perform forward simulation in the inferred model by examining how changes in the rates affect the emergent animal behaviors.

Our approach is to be contrasted with standard approaches which postulate constant (effective, or average) behavior in time. While that single behavior can be complex, e.g., it can represent a complex computation to determine an animal's movement based on the position of its neighbors, the computation is usually assumed to be static through time and (most often) also deterministic. Our model can be seen as an extension of this standard approach to multiple behaviors (e.g., motion computations), possibly very different from each other, that the animal can stochastically switch between, in a way that is influenced by external and internal variables. While this is relevant for describing motion through space---especially to accurately capture abrupt changes in velocity or direction---there also exist many behaviors of interest that play out in physical space but extend beyond motion computation. Such behaviors often consist of stereotyped, discrete events (body pose changes during courtship dances, vocalizations, tailbeats, grooming behaviors, etc.) that are poorly described by continuous, averaged response models. With a proper choice of behavioral states, our model has the ability to capture this broad range of biological phenomena, either at the level of an isolated individual moving in space or interacting within a group.

\section*{Acknowledgments}
This work was supported by the Human Frontiers Science Foundation (HFSP) grant to GT and ES.

\appendix
\setcounter{table}{0}
\setcounter{figure}{0}
\numberwithin{equation}{section}
\numberwithin{figure}{section}

\section{Impact of sampling frequency} \label{ssec_sampling}
Experimental trajectories contain measurement noise but also a finite resolution due to the minimal sampling frequency that is practical to use in the experiments. If the sampling frequency is very large, a long enough trajectory contains enough information to resolve the transition rates accurately. However, if the sampling frequency is not large, even if all transitions are captured, the inference may be inaccurate due to a low resolution of the integral in \eqref{likelihood} (or similarly the second term in \eqref{likelihood_hist}). 
\begin{figure}[h]
	\centering
	\includegraphics[width=0.7\textwidth]{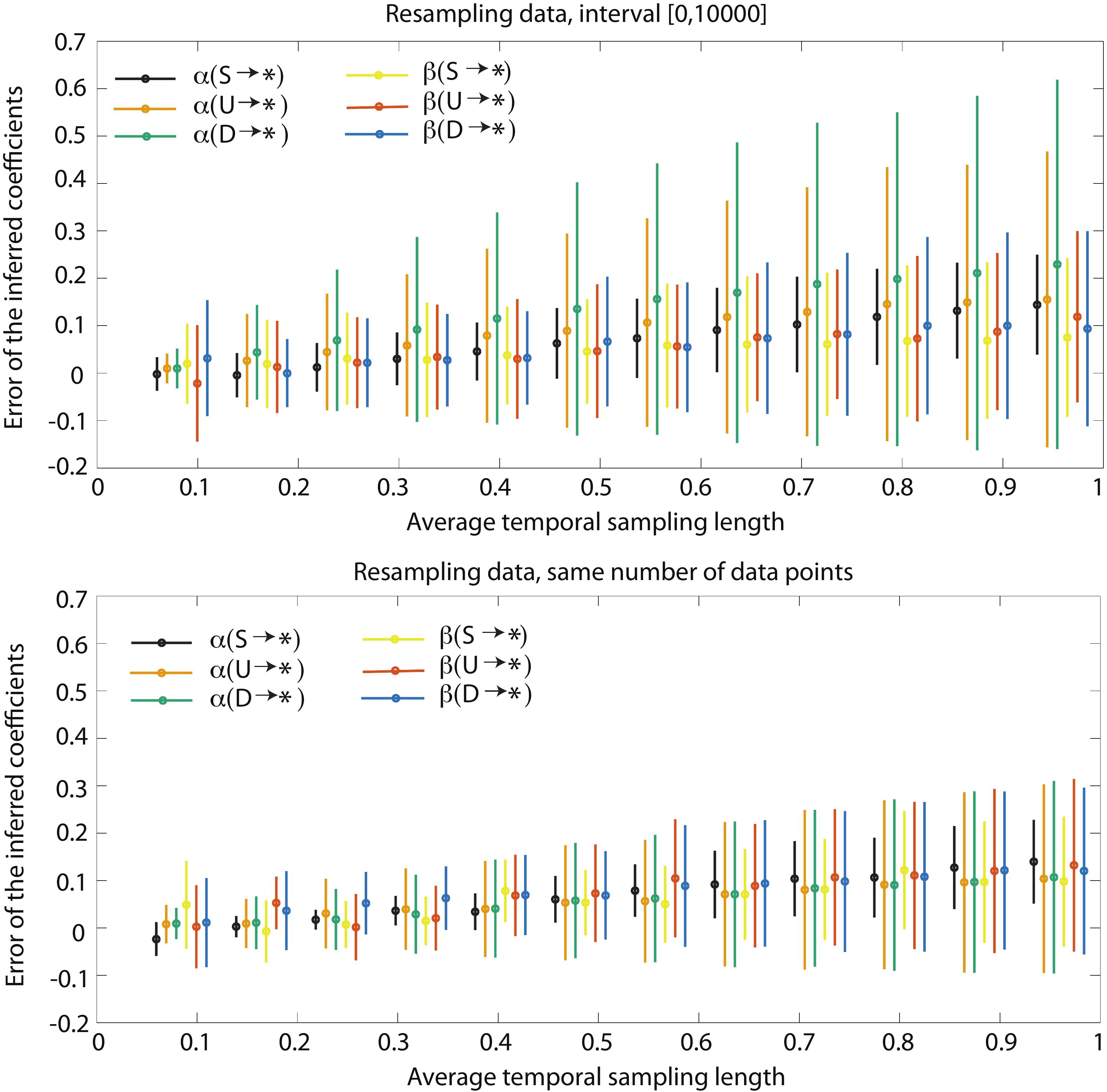}
	\caption{Accuracy of the inference on sampling frequency. Two interacting ants are simulated according to 
	the same model as in Figure~\ref{fig_2ants} for $t\in[0,10\,000]$. The numerical integration of the 
	dynamical equations uses an adaptive time step with a mean $\langle \Delta t \rangle = 0.0795$. The data are re-sampled by taking
	every $k$-th datapoint, for the purpose of error sensitivity analysis. The error is defined as the mean difference between the real and inferred coefficient, 
	$\langle \hat \alpha(x \ra \ast) - \alpha(x \ra \ast) \rangle$ (analogous for $\beta$). The average is taken only for   $z, dz \in [-2,2]$, where enough data are available.
	(region of no regularization). (A) Error in coefficients $\alpha$ and $\beta$ inferred from the simulation in
	$t\in[0,10\,000]$ and different sampling rates, $\langle \Delta t\rangle = 0.0795 k$. 
	(B) Error in coefficients $\alpha$ and $\beta$ inferred from 
	the simulation in	$t\in[0,10\,000k]$ and different sampling rates, $\langle \Delta t \rangle = 0.0795k$.
	}
	\label{fig_sampling}
\end{figure}

It is important to understand how this sampling frequency affects the quality of the inference. To test this we chose two interacting ants as a toy example and simulated the stochastic dynamics for a fixed period of time $t\in[0,10\,000]$. We then resampled the trajectory by taking only every $k$-th datapoint into consideration. Since the simulation ran with an adaptive time step we report only the average time step throughout the full simulation, $\langle \Delta t \rangle = 0.0795$. The results shown in Figure~\ref{fig_sampling} (A) compare the accuracy of the inferred transition rates as a function of the average sampling step for data in $t\in[0,10\,000]$. It is important to note that resampling does not always reduce the number of observed transitions but it reduces the number of data points between them. 

As expected, the inference error grows as the sampling frequency decreases. But is this phenomenon simply a reflection of the decreased number of the data points? We study this by running a few sets of stochastic simulations with different sampling frequencies but with the same number of datapoints in total. This is done by running the coarser trajectories for a longer time, time intervals used are from $[0,10^4]$ to $[0,12\times10^4]$. While the top panel in Figure~\ref{fig_sampling} shows a loss of accuracy due to both the smaller time coverage density and due to a smaller number of datapoints, Figure~\ref{fig_sampling}(B) keeps the number of datapoints constant thus reflecting only the density of data points. Despite the inference is done with the same number of data points  the error still grows in time at a linear rate with the decrease in the sampling frequency.

\section{Coarse-graining through stochastic states with constant transition rates} \label{ssec_CG}

Here we study an example of a hybrid system with a subset of behavioral states, which are not distinguishable in the experiment. We refer to those as microscopic states, whereas the states that are observable are macroscopic by definition. Ideally, we want to infer the rates of the full model (macro + micro) given only information about the macroscopic system. For that to be possible we assume that 
\begin{itemize}
	\item 
	All microscopic states are identical, i.e., the transition from any microstate to a given accesible macrostate is the same. 
	Similarly, the transition from a macrostate to any allowed microstate is the same.
	\item 
	The transitions between the microstates are constant and the same.
	\item 
	The microstates may follow different laws of motion.
\end{itemize}
The idea is then to replace the set of identical microstates  by a summary state which is macroscopic. However, this state is no longer deterministic; the laws of motion are randomly switching between the laws of motion of the corresponding microstates. Thus such an approach is an extension of the hybrid model as described in Section~\ref{ssec_models}. 

We chose the tumble-and run motion as a prototype for this demonstration. This is because the tumble left ($T\!L$) and tumble right ($T\!R$) form a great example of the unobserved internal states that can be clumped together to an observed tumble ($T$) state. The run ($R$) state does not contain any unobserved microstates and the transitions between the $T$ and $R$ states are experimentally observable.

Thus, the two levels of the tumble and run model are: the macroscopic level with two internal states $S^M = \{ {T, R}\}$ (both experimentally observable), and the microscopic level with two microstates: $S^m = \{ T\!L, T\!R \}$. The two models differ in replacing  two microstates (tumble left and tumble right) by a summary tumble state. The $T\!L$ and $T\!R$ states follow different laws of motion but the transitions between them are assumed to be constant, i.e., independent of the position and angle $(x,y,\ph)$.

First note that we can compute the probability of the realized microscopic path \eqref{eq_path} using all transition points ${t_k}$ but we can also compute a coarse-grained version using just the transition points ${t_{k'}}$, where the set of microscopic transition indices $k$ is $\{1,2,\dots,K\}$ and the set of the macroscopic indices is a subset of it. Such a coarse-grained expression leads to a log-likelihood that involves only the transitions among the observable states. To infer these macroscopic rates one needs to have access to a trajectory $\mathbf z(t)$ during the tumble state. This is an unobserved stochastic process due to random transitions between the left/right motion and thus we will seek a statistical description of it in terms of the probability density function of a position $h(\mathbf z,t)$ after elapsed time $t$ from the transition into the tumble state.

The problem can be formulated as a set of the coupled advection-reaction equations for the probability densities $u(\ph,t)$ and $v(\ph,t)$ that the system which entered the composite tumble state at time $t=0$ is in the tumble left and right state at time $t$, respectively with an angle $\ph$. 
The system has a form
\begin{eqnarray}
	u_t &=& -\omega u_\ph - \gamma_1 u + \gamma_2 v\,, \label{PDEu}\\
	v_t &=& \phantom{-}\omega v_\ph + \gamma_1 u - \gamma_2 v\, \label{PDEv}. 
\end{eqnarray}
where the reaction terms capture changes of state at the constant rate $\gamma$ and the advection terms capture the change in angle due to the motion with angular velocity $\omega$.
The system is defined for $\ph \in[0,2\pi]$ is complemented by the periodic boundary conditions 
\begin{eqnarray}
	&u(0,t) = u(2\pi,t)\,,  &u_\ph (0,t) = u_\ph (2\pi,t) \label{BCu}\,,\\
	&v(0,t) = v(2\pi,t)\,, &v_\ph (0,t) = v_\ph (2\pi,t) \label{BCv}\,.
\end{eqnarray}
At the transition into the tumble state the position and the angle of the individual are $(x_0,y_0,\ph_0)$. We assume that the left/right states are initially equally likely, therefore 
\begin{eqnarray}
	u(\ph,0) = \frac12 \delta(\ph-\ph_0)\,, \quad v(\ph,0) = \frac12 \delta(\ph-\ph_0)\,. \label{ICuv}
\end{eqnarray}
Note that due to translational symmetry the problem needs to be solved only for one value of $\ph_0$ since 
\begin{eqnarray}
	u(\ph,\ph_0,\omega,\alpha,t) = u(\ph+\ph_0,0,\omega,\alpha,t) \label{translation}
\end{eqnarray} 
and similar for $v$ (where we included dependence on all the parameters). The above process is called a velocity jump process and it has been studied in a mathematical biology literature using the telegrapher's equation \cite{goldstein1951,kac1974,othmer1988}.

\begin{figure}[h]
	\centering
	\includegraphics[width=0.6\textwidth]{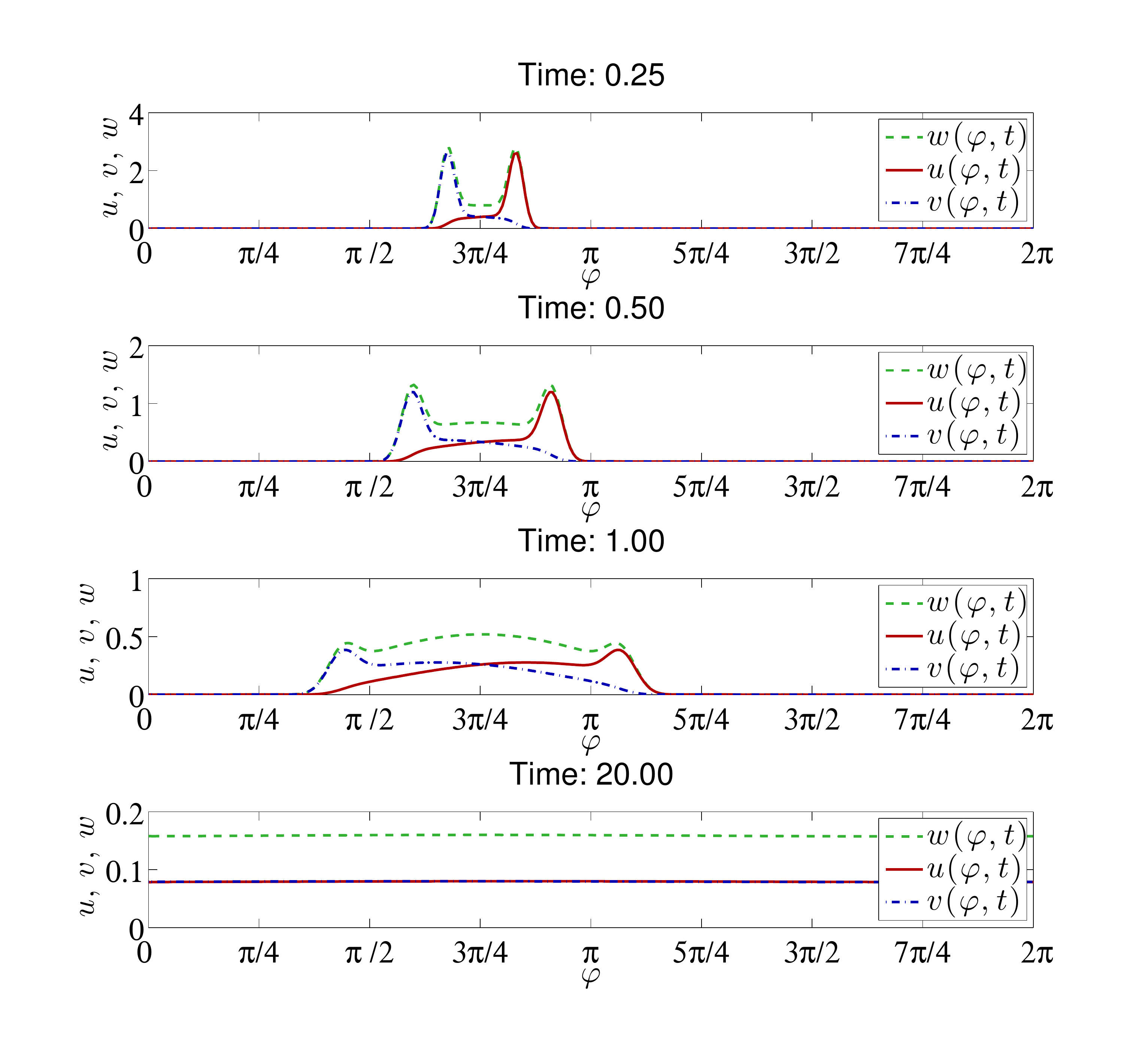}
	\caption{
	The solution of the system \eqref{PDEu}-\eqref{PDEv} at times $t=0.25,0.5,1, 20$ for 
	parameters $\omega = 1$, $\gamma = 2$, $\ph_0 = 3\pi/4$.
	We have used an upwind method that replaces the spatial derivatives of $u$ and $v$ using a backward 
	and forward finite differences, respectively and the temporal derivative by a forward difference. 
	This method is stable for $\Delta t > \frac{\Delta \ph}{\omega} $ that we ensure by choosing 
	$\Delta t = 2 \frac{\Delta \ph}{\omega}$.
	}
	\label{fig_PDEsol}
\end{figure}

Figure~\ref{fig_PDEsol} shows the solution of the coupled PDE system at times $t=0.25,0.5,1, 20$. The function $u(\varphi,t)$ and $v(\varphi,t)$ combine into a density function $w = u + v$ for the orientation at time $t$. For short times $t \ll \omega/\gamma$ the system carries information about the entrance angle and shorter transitions times give a rise to a positive correlation between the entrance and the exit angles. This correlation disappears when $t\gg \omega/\gamma$ where the exit distributions $u$ and $v$ become close to uniform distributions.

Once the solution of \eqref{PDEu}-\eqref{PDEv} is obtained we may formulate the inference problem using the coarse-grained log-likelihood function with an uniform prior
\begin{eqnarray}
	L(\boldsymbol \alpha) &=& \sum_{ijk} \sum_{x,x'\in S^{M}}
	\alpha_{ijk}( x \ra  x', \mathbf z_{ijk}) n( x \ra x', \mathbf z_{ijk})  \nonumber\\
	&+&  \sum_{ijk} \sum_{x,x'\in S^{M}} \exp \left ( \alpha_{ijk}(x \ra  x', \mathbf z_{ijk}) \right) T( x, \mathbf z_{ijk}) 
	+C \label{CGlikelihood}
\end{eqnarray}
where $\mathbf z_{ijk} = (x_i,y_j,\ph_k)$ is the center of the $ijk$-th bin in the tiling functions 
and $m$ is the number of the divisions of the interval of interest (in Figure~\ref{fig_RT} we used $x\in[-10,30]$, $y\in [-10,10]$, $\ph\in[0,2\pi]$ and $m=20$).

Observing the system's macroscopic states $\{x\} \in S^M$ and positions $\{ \mathbf z\}$ at transition times leads to the statistics $n(x\ra x',\mathbf z)$. To obtain the statistics $T(x,\mathbf z)$ one needs to approximate the amount of time the organism has spent at each bin of the domain. This can be calculated from a list of all entrance angles into the tumble state and the corresponding duration before leaving it to the run state. For each event from this list we compute numerically the density function of the angle $\ph$ after time $t$ spent in the coarse-grained macrostate ($T$) with the entrance value $\ph_0^k$ and the microscopic transition rates $\gamma$
\begin{eqnarray}
	w(\ph,\ph_0,\omega,\gamma,t) = 
	u(\ph,\ph_0,\omega,\gamma,t)+v(\ph,\ph_0,\omega,\gamma,t) \label{dist_micro}
\end{eqnarray}
for $t\in[0,\tau_k]$  and use it as a proxy of how much time the system spends at a given $\ph$-bin at each time. Note that the parametric dependence on $\ph_0$ can be suppressed due to translational symmetry \eqref{translation} and the answer depends only on the parameters $\gamma$ and $\omega$, thus the solution of  \eqref{PDEu}-\eqref{PDEv} has to be computed just once. For the given values $\gamma, \omega$ we use the entrance angles $\{\varphi_0\}$  into the macrostate $T$ and the time spent in this state $\{\tau\}$ to sample the statistics $T(x,\mathbf z)$ for the state $T$. For  every bin $D_{ijk}$ and a chosen time step $\Delta t$ we evaluate the sum
\begin{eqnarray}
	T_{ijk}(x,\mathbf z|\gamma,\omega) = \chi(\mathbf z \in D_{ijk} ) \sum_n \sum_{m=0}^{M} \chi(\tau_n > (m+1)\Delta t) w(\ph+\ph_0^n,0,\omega,\gamma,m\Delta t) \Delta t\,, \label{Tdistribution}
\end{eqnarray}
composed of summands for every pair $(\ph_n,\tau_n)$ and adding contributions at time $t\in[0,\tau]$, discretized to $\Delta t$-sized intervals. The function $\chi$ is a characteristic function, that results in one if the argument is true and zero otherwise. 

As long as we have enough data for sampling the bulk part of the distribution $T(x,\mathbf z|\gamma,\omega)$ the coarse-grained log-likelihood will be close to the actual one. In practice, the parameters $\gamma$ and $\omega$ are not known. Therefore we formulate a composite log-likelihood including also the microscopic parameters  $\gamma, \omega$ and find the maximum of a function of macroscopic parameters $\boldsymbol \alpha^M$, appended by  $\gamma,\omega$
\begin{eqnarray}
	L(\boldsymbol \alpha^M,\gamma,\omega) &=&
	\sum_{ijk} \sum_{x,x'\in S^M}
	\alpha_{ijk}( x \ra  x', \mathbf z_{ijk}) n( x \ra x', \mathbf z_{ijk}) \\
	&+& 
	\sum_{ijk} \sum_{x,x'\in S^M}
	\exp \left ( \alpha_{ijk}(x \ra  x', \mathbf z_{ijk}) \right) T( x, \mathbf z_{ijk}) 
	\label{CGlikelihood2}\,.
\end{eqnarray}
The implementation of the stochastic state approach for the bacterial chemotaxis can be found in the main text in Section~\ref{ssec_bacteria}.

\section{Simulation code}
To explore further details of our method readers can access our implementation in {\it Matlab} which is part of the supplemental information.  It contains two toy examples: simplified motion of ants in 1D and bacteria in a chemical gradient in 2D. The code contains stochastic simulator of the motion of the organisms, assuming known transition rates. The simulated trajectories are then used as an input for the inference problem where the transition rates are reconstructed. We used a decomposition of the signal using tiling functions $\ph$, $\Psi$ exclusively. The single ant inference performs optimization in a bin-by-bin matter, while for multiple interacting ants inference is global and uses the whole trajectories. The bacterial chemotaxis Section~\ref{ssec_bacteria} is implemented on two scales: microscopic where all states are assumed observable; and macroscopic where only the states tumble and run are observed. More detail can be found in Appendix~\ref{ssec_CG}.

We use a library {\it minFunc} containing the L-BFGS method to compute the optimum of the log-likelihood since this is a recommended method for large parameter unconstrained optimization in the information theory. To obtain good enough approximation of the transition rates follow the numbers in Sections~\ref{ssec_1ant} and \ref{ssec_bacteria}.


\end{document}